\documentclass[prd,twocolumn,showpacs,amsmath,amssymb,letterpaper]{revtex4}
\usepackage{graphicx}
\usepackage{bm}
\usepackage{epsf}
\usepackage{color}

\begin{document}

\title{Viability of $\Delta m^2\sim$~1 eV$^2$ sterile neutrino mixing models in light of MiniBooNE electron neutrino and antineutrino data from the Booster and NuMI beamlines}

\author{G.~Karagiorgi$^1$}
\email{georgiak@mit.edu}
\author{Z.~Djurcic$^2$}
\email{zdjurcic@nevis.columbia.edu}
\author{J.~M.~Conrad$^1$}
\email{conrad@mit.edu}
\author{M.~H.~Shaevitz$^2$}
\email{shaevitz@nevis.columbia.edu}
\author{M.~Sorel$^3$\bigskip}
\email{sorel@ific.uv.es}

\affiliation{$^1$Massachusetts Institute of Technology, Cambridge, MA 02139}
\affiliation{$^2$Department of Physics, Columbia University, New York, NY 10027}
\affiliation{$^3$Instituto de F\'{i}sica Corpuscular, IFIC, CSIC and Universidad de Valencia, Spain\bigskip}

\smallskip

\date{\today}

\begin{abstract}
This paper examines sterile neutrino oscillation models in light of recently published results from the MiniBooNE Experiment. The new MiniBooNE data include the updated neutrino appearance results, including the low energy region, and the first antineutrino appearance results, as well as first results from the off-axis NuMI beam observed in the MiniBooNE detector. These new global fits also include data from LSND, KARMEN, NOMAD, Bugey, CHOOZ, CCFR84, and CDHS. Constraints from atmospheric oscillation data have been imposed.  We test the validity of the three-active plus one-sterile (3+1) and two-sterile (3+2) oscillation hypotheses, and we estimate the allowed range of fundamental neutrino oscillation parameters in each case. We assume CPT-invariance throughout. However, in the case of (3+2) oscillations, CP violation is allowed. With the addition of the new MiniBooNE data sets, there are clear incompatibilities between neutrino and antineutrino experiments under a (3+1) oscillation hypothesis. A better description of all short-baseline data over a (3+1) is provided by a (3+2) oscillation hypothesis with CP violation. However, we still find large incompatibilities among appearance and disappearance experiments, consistent with previous analyses, as well as incompatibilities between neutrino and antineutrino experiments. Aside from LSND, the data sets responsible for this tension are the MiniBooNE neutrino data set, CDHS, and the atmospheric constraints. On the other hand, fits to antineutrino-only data sets, including appearance and disappearance experiments, are found significantly more compatible, even within a (3+1) oscillation scenario. 
\end{abstract}


\maketitle

\section{\label{sec:one}INTRODUCTION}

Sterile neutrino oscillation models were proposed more than a decade ago as an explanation for the LSND anomaly \cite{lsnd,Peres:2000ic,Strumia:2002fw,Grimus:2001mn,Sorel:2003hf}, an excess of events consistent with $\bar \nu_\mu \rightarrow \bar \nu_e$ oscillations at high $\Delta m^2$. These models relate $\nu_e$ appearance ($\nu_{\mu}\rightarrow\nu_e$) with $\nu_\mu$ and $\nu_e$ disappearance ($\nu_{\mu}\rightarrow\nu_{\not{\mu}}$ and $\nu_e\rightarrow\nu_{\not{e}}$), motivating combined fits in all three oscillation channels. Relatively early in the discussion of models, it was demonstrated \cite{Sorel:2003hf,Maltoni:2002xd} that a three-active plus one-sterile (3+1) neutrino oscillation model could not reconcile the LSND result with existing null results from other short-baseline (SBL) experiments, including KARMEN \cite{Armbruster:2002mp}, NOMAD \cite{nomad}, Bugey \cite{Declais:1994su}, CHOOZ \cite{Apollonio:2002gd}, CCFR84 \cite{Stockdale:1984cg}, and CDHS \cite{Dydak:1983zq}, which had similar high $\Delta m^2$ sensitivity.  However, it was shown that a three-active plus two-sterile neutrino (3+2) oscillation scenario provided a better description of these data sets \cite{Sorel:2003hf}. 

\indent In 2001, the MiniBooNE experiment began running with the goal to test the LSND result using both neutrino and antineutrino beams.  This is a short-baseline appearance and disappearance experiment located at Fermi National Accelerator Laboratory (Fermilab). MiniBooNE's first results, reported in 2007, described a search for $\nu_{\mu}\to\nu_e$ oscillations \cite{mbnu}. These results were incompatible with a simple two-neutrino oscillation interpretation of the LSND signal and, within this model, MiniBooNE excluded the LSND result at the 98\% CL.  However, this same analysis reported a 3.7$\sigma$ excess of electron neutrino candidate events at low energies, between 300-475 MeV, which remains unexplained. Reference \cite{Maltoni:2007zf} included the MiniBooNE first result in a global fit to all SBL experiments under the (3+1) and (3+2) oscillation scenarios.  The analysis built on an earlier study, which introduced the possibility of CP violation $(P(\nu_\mu \rightarrow \nu_e) \ne P(\bar \nu_\mu \rightarrow \bar \nu_e))$ within (3+2) fits \cite{karagiorgi}.  Including the first MiniBooNE results into the global fit led to two observations in Ref.~\cite{Maltoni:2007zf}: 1) MiniBooNE, LSND, and the null appearance experiments (KARMEN and NOMAD) are compatible under a (3+2) sterile neutrino oscillation scenario with large CP violation.  2) There is severe tension between appearance and disappearance experiments, at a level of more than 3$\sigma$.  In this paper we will consider both observations in light of new appearance data. Also, we will show that the incompatibility between appearance and disappearance experiments arises mainly from two $\nu_{\mu}$ disappearance data sets: CDHS and atmospheric constraints. 

\indent Motivated by three new results from MiniBooNE, this paper re-examines the (3+1) and (3+2) global fits to the SBL data.  These new results are: 1) an updated $\nu_\mu \rightarrow \nu_e$ result \cite{mblowe}; 2) first results for a $\bar{\nu}_{\mu}\to\bar{\nu}_e$ search \cite{mbnubar}; and 3) first $\nu_{\mu}\to\nu_e$ results from the NuMI off-axis beam at MiniBooNE \cite{numi}.  We consider these new results in combination with seven SBL data sets.  These provide constraints on: $\nu_{\mu}$ disappearance (from the CCFR84 and CDHS experiments), $\bar{\nu}_e$ disappearance (from the Bugey and CHOOZ experiments), $\nu_{\mu}\to\nu_e$ oscillations (from the NOMAD experiment), and $\bar{\nu}_{\mu}\to\bar{\nu}_e$ oscillations (from the LSND and KARMEN experiments). Furthermore, we have taken into account atmospheric constraints based on the analysis of Ref.~\cite{Maltoni:2004gf}. These constraints have been incorporated in our analysis following the method described in Ref.~\cite{karagiorgi}, and are included in fits to all SBL experiments, null SBL experiments, or as explicitly stated. Table \ref{tab:sblsummary} summarizes all SBL data sets used in the fits presented in this paper.

\begingroup
\squeezetable
\begin{table}[tb] 
\vspace{0.2cm}
\begin{ruledtabular} 
\begin{tabular}{lc} 
Data Set & Channel  \\ \hline\hline
{\bf Appearance experiments:} & \\
LSND                            & $\bar{\nu}_{\mu}\to\bar{\nu}_e$ \\
BNB-MB($\nu$)                   & $\nu_{\mu}\to\nu_e$   \\
BNB-MB($\bar{\nu}$)             & $\bar{\nu}_{\mu}\to\bar{\nu}_e$  \\
NUMI-MB                         & $\nu_{\mu}\to\nu_e$   \\
KARMEN                          & $\bar{\nu}_{\mu}\to\bar{\nu}_e$  \\
NOMAD                           & $\nu_{\mu}\to\nu_e$   \\
\hline
{\bf Disappearance experiments:} & \\
Bugey                           & $\bar{\nu}_e\to\bar{\nu}_{\not{e}}$  \\
CHOOZ                           & $\bar{\nu}_e\to\bar{\nu}_{\not{e}}$  \\
CCFR84                          & $\nu_{\mu}\to\nu_{\not{\mu}}$  \\
CDHS                            & $\nu_{\mu}\to\nu_{\not{\mu}}$  \\
\end{tabular} 
\end{ruledtabular} 
\caption{\label{tab:sblsummary} Short-baseline oscillation data sets considered in this paper, and oscillation channel that each data set constrains.}
\end{table} 
\endgroup

\indent In this work, we do not discuss experimental constraints on sterile neutrino models other than SBL and atmospheric neutrino ones. Constraints from the measurement of the electron spectrum near the endpoint in beta-decay experiments are relatively weak as long as the mostly-sterile mass states are heavier than the mostly-active ones, because of the small electron flavor of the former (see Refs.~\cite{Sorel:2003hf,Goswami:2007kv}). We make this assumption throughout the paper, by requiring that the heavier sterile neutrino mass eigenstates, $m_5$ and $m_4$, obey $m_5 > m_4 > m_1$. Constraints on the energy density (and mass) in the Universe carried by sterile neutrinos from cosmic microwave background, matter power spectrum, and supernovae data have been studied in Ref.~\cite{Melchiorri:2008gq}. While relevant, these constraints are found to be weaker than SBL ones, since sterile neutrino states do not necessarily feature thermal abundances at decoupling. Constraints on the number of relativistic degrees of freedom from the observations of cosmological abundances of light elements produced at the epoch of Big Bang Nucleosynthesis may also be relevant, although model-dependent. For such a study involving one sterile neutrino species participating in the mixing, see for example Ref.~\cite{Dolgov:2003sg}.

\indent The paper is organized as follows. In Section \ref{sec:mbresults}, we provide a short description of the MiniBooNE experiment and the new data sets.  In Section \ref{sec:two}, we specify the formalism used in this analysis to describe (3+$n$) oscillations, where $n$ is the number of sterile neutrinos. In Section \ref{sec:three}, we discuss the analysis method followed, and describe in detail the way in which the three MiniBooNE data sets have been incorporated. In Section \ref{sec:four}, we present the results obtained for the (3+1) (CP-conserving only), and (3+2) CP-conserving and CP-violating hypotheses. For each hypothesis, we quote the compatibility between various sets of SBL experiments and report the best-fit neutrino mass and mixing parameters derived from the combined analysis of all experimental data sets. In the (3+2) CP-violating case, we discuss the constraints on the CP violation phase, inferred from a combined analysis of all SBL oscillation results. Finally, in Section \ref{sec:five}, we discuss constraints to the (3+2) CP-violating models from each of the SBL experiments considered in this analysis. The goal of this particular study was to investigate whether the source of tension between appearance and disappearance experiments \cite{Maltoni:2007zf} is a result of a single experiment, other than LSND.


\section{The New MiniBooNE Data Sets \label{sec:mbresults}}

The MiniBooNE experiment uses a muon neutrino beam produced by 8 GeV protons from the Fermilab Booster Neutrino Beamline (BNB) impinging on a beryllium target. The target is located within a magnetic focusing horn \cite{mbflux}. The current of the horn can be reversed for running neutrinos or antineutrinos, allowing MiniBooNE to perform both neutrino and antineutrino oscillation searches. The detector \cite{mbdet} is located $L=541$~m from the primary target, and the neutrino flux has an average energy of $\sim 0.75$ GeV. This design maintains the LSND $L/E$ of $\sim 1$~m/MeV. The detector consists of a spherical tank with a 610-cm active radius, instrumented with 1520 8-inch photomultipliers. This is filled with 800 tons of pure mineral oil. An outer veto region rejects cosmic rays and neutrino events producing particles which cross the detector boundaries.

\indent The MiniBooNE neutrino data set used in this analysis corresponds to the updated results recently reported by the MiniBooNE collaboration \cite{mblowe}. Compared to the first MiniBooNE result which was released in 2007 \cite{mbnu}, the new result involves a re-analysis of the MiniBooNE low energy excess events and several updates to the Monte Carlo prediction. These updates include a new model of photonuclear effects, incorporation of new data on $\pi^0$ production and a better treatment of pion re-interaction in the detector following decay, an improved estimate and rejection method of the background from interactions outside the detector, and improvements to the determination of systematic errors. The updated low-energy analysis has resulted in a reduction to the significance of the excess from 3.7$\sigma$ in the original analysis to 3.4$\sigma$, along with some slight modification to the shape of the energy spectrum; specifically, the peak of the excess has shifted slightly to higher neutrino energies. In addition, the new analysis extends in energy down to 200 MeV, compared to 300 MeV in the original analysis, which offers additional $L/E$ information.  The new result also corresponds to modestly higher statistics, corresponding to the total data collected during the experiment's neutrino running of 6.46$\times$10$^{20}$ protons on target (POT), compared to 5.58$\times$10$^{20}$, previously. 

\indent More recently, the MiniBooNE Collaboration reported its first results from a search for $\bar{\nu}_{\mu}\to\bar{\nu}_e$ oscillations, using a muon antineutrino beam \cite{mbnubar}. The antineutrino analysis performed by MiniBooNE mirrors the updated neutrino analysis \cite{mblowe}, and includes the Monte Carlo prediction updates of the latter. The total antineutrino data set used in the analysis corresponds to 3.39$\times$10$^{20}$ POT. However, due to meson production and cross-section effects, the antineutrino event sample, unlike the neutrino event sample, is statistically limited. Unlike the neutrino search, the MiniBooNE antineutrino search provides a direct test of the LSND result, similar to the search performed by KARMEN. The MiniBooNE sensitivity to $\bar{\nu}_{\mu}\to\bar{\nu}_e$ extends into the low-$\Delta m^2$ region allowed by a combined analysis of KARMEN and LSND data.  Nevertheless, the MiniBooNE antineutrino search has observed no conclusive signal, and a limit has been set, which is considerably weaker than the sensitivity, and comparable to the KARMEN limit. The limit degradation with respect to the sensitivity is due to a 2.8$\sigma$ fluctuation of data above expected background observed in the 475-675 MeV energy region. Thus, at present, the MiniBooNE antineutrino result is inconclusive with respect to oscillations allowed by LSND. However, MiniBooNE is in the process of collecting more antineutrino data. This is expected to improve the experiment's sensitivity to $\bar{\nu}_{\mu}\to\bar{\nu}_e$ oscillations.  Updated results are expected after about three years of running. 

\indent The third new data set \cite{numi} arises from the fact that the MiniBooNE detector is illuminated by the off-axis (110 mrad) neutrino flux from the NuMI beamline at Fermilab. This analysis has reported a 1.2$\sigma$ excess of $\nu_e$-like events in the neutrino energy range below 900 MeV. The NuMI data set corresponds to a mean $L/E$ that is approximately the same as those of the MiniBooNE and LSND data sets, and therefore probes the same $\Delta m^2$ range, providing complementary information in oscillation fits with MiniBooNE and LSND.


\section{\label{sec:two}(3+$n$) STERILE NEUTRINO OSCILLATION FORMALISM}

The formalism used in this paper follows that which was presented in Ref.~\cite{karagiorgi}.  We provide a brief summary here.

\indent In sterile neutrino oscillation models, under the assumptions of CPT invariance and negligible matter effects, the probability for a neutrino produced with flavor $\alpha$ and energy $E$, to be detected as a neutrino of flavor $\beta$ after traveling a distance $L$, is given by \cite{Barger:1999hi,Kayser:2002qs}:
\begin{eqnarray}
\label{eq:oscprob}
P(\nu_{\alpha}\to\nu_{\beta})= & \delta_{\alpha\beta}-4\sum_{i>j}\mathcal{R}(U^{\ast}_{\alpha i}
 U_{\beta i}U_{\alpha j}U^{\ast}_{\beta j})\sin^2x_{ij}+\nonumber \\
 & 2\sum_{i>j}\mathcal{I}(U^{\ast}_{\alpha i}U_{\beta i}U_{\alpha j}U^{\ast}_{\beta j})\sin2x_{ij}
\end{eqnarray}
\noindent where \begin{math}\mathcal{R}\end{math} and \begin{math}\mathcal{I}\end{math} indicate the real and imaginary parts of the product of mixing matrix elements, respectively; $\alpha,\beta\equiv e,\mu ,\tau$, or $s$, ($s$ being the sterile flavor); $i,j=1,\ldots ,3+n$ ($n$ being the number of sterile neutrino species); and $x_{ij}\equiv 1.27\Delta m^2_{ij}L/E$. In defining $x_{ij}$, we take the neutrino mass splitting $\Delta m^2_{ij}\equiv m^2_i-m^2_j$ in $\hbox{eV}^2$, the neutrino baseline $L$ in km, and the neutrino energy $E$ in GeV.  For antineutrinos, the oscillation probability is obtained from Eq.~\ref{eq:oscprob} by replacing the mixing matrix $U$ with its complex-conjugate matrix. Therefore, if the mixing matrix is not real, neutrino and antineutrino oscillation probabilities can differ.

\indent For 3+$n$ neutrino species, there are, in general, $2+n$ independent mass splittings, $(3+n)(2+n)/2$ independent moduli of parameters in the unitary mixing matrix, and $(2+n)(1+n)/2$ Dirac CP-violating phases that may be observed in oscillations. In SBL neutrino experiments that are sensitive only to $\nu_{\mu}\to\nu_{\not{\mu}}$, $\nu_e\to\nu_{\not{e}}$, and $\nu_{\mu}\to\nu_e$ transitions, the set of observable parameters is reduced considerably.  In this case, the number of observable parameters is restricted to $n$ independent mass splittings, $2n$ moduli of mixing matrix parameters, and $n-1$ CP-violating phases. Therefore, for (3+2) sterile neutrino models ($n=2$ case), for example, there are two independent mass splittings, $\Delta m^2_{41}$ and $\Delta m^2_{51}$, both defined to be greater than zero, four moduli of mixing matrix parameters $|U_{e4}|,\ |U_{\mu 4}|,\ |U_{e5}|,\ |U_{\mu 5}|$, and one CP-violating phase. The convention used for the CP-phase is:
\begin{equation}
\label{eq:cpvphase}
\phi_{45}=arg(U_{\mu 5}^*U_{e5}U_{\mu 4}U_{e4}^* ).
\end{equation}
\noindent In that case, the general oscillation formula in Eq.~\ref{eq:oscprob} becomes:
\begin{eqnarray}
\label{eq:threeplustwo_1}
P(\nu_{\alpha}\to\nu_{\alpha}) = 1-4[(1-|U_{\alpha 4}|^2-|U_{\alpha 5}|^2)\cdot \nonumber\\
(|U_{\alpha 4}|^2\sin^2 x_{41}+|U_{\alpha 5}|^2\sin^2 x_{51})+ \nonumber \\
|U_{\alpha 4}|^2|U_{\alpha 5}|^2\sin^2 x_{54}] 
\end{eqnarray}
\noindent and
\begin{eqnarray}
\label{eq:threeplustwo_2}
P(\nu_{\alpha}\to\nu_{\beta\ne\alpha}) = 4|U_{\alpha 4}|^2|U_{\beta 4}|^2\sin^2 x_{41}+ \nonumber \\
      4|U_{\alpha 5}|^2|U_{\beta 5}|^2\sin^2 x_{51}+ \nonumber \\
      8|U_{\alpha 5}||U_{\beta 5}||U_{\alpha 4}||U_{\beta 4}|
 \sin x_{41}\sin x_{51}\cos (x_{54}-\phi_{45}) 
\end{eqnarray}
\indent The formulas for antineutrino oscillations are obtained by substituting $\phi_{45}\to -\phi_{45}$.

\indent For the case of (3+1) sterile neutrino models ($n=1$ case), the corresponding oscillation probabilities are obtained from Eqs.~\ref{eq:threeplustwo_1} and \ref{eq:threeplustwo_2} by setting $x_{51}=x_{54}=0$ and $|U_{\alpha 5}|=0$. Note that, under the above assumptions, no CP violation is allowed for (3+1) models.


\section{\label{sec:three}ANALYSIS METHOD}

In this section, we first provide an overview of the fitting technique. We then focus on the method followed for including the MiniBooNE data sets. The physics- and statistical-assumptions for the other null SBL experiments and LSND, which are also included in the fit, are described in detail in Ref.~\cite{Sorel:2003hf}. The constraints from atmospheric experiments, according to Ref.~\cite{Maltoni:2004gf}, have been incorporated as described in Ref.~\cite{karagiorgi}. 

\subsection{\label{sec:three0}General Technique}

\indent The Monte Carlo method used to apply the oscillation formalism described in Section \ref{sec:two} closely follows the one described in Ref.~\cite{karagiorgi}. We start by randomly varying sets of oscillation parameters: $\Delta m^2_{41}, |U_{e4}|, |U_{\mu 4}|$ for the case of (3+1); $\Delta m^2_{41}, |U_{e4}|, |U_{\mu 4}|, \Delta m^2_{51}, |U_{e5}|, |U_{\mu 5}|, \phi_{45}$ for the case of (3+2).  Without loss of generality, we take $\Delta m^2_{51}>\Delta m^2_{41}$. In CP-conserving models, $\phi_{45}$ is set to 0 or $\pi$ by default, whereas in CP-violating models $\phi_{45}$ is allowed to vary within the full $(0,2\pi)$ range.  For each set of oscillation parameters, a signal prediction is obtained and compared to observed data for each SBL experiment, in the form of a $\chi^2$ for each experiment.  For each set of oscillation parameters that is generated, the various $\chi^2$'s are linearly summed together to form $\chi^2_{SBL}$, which is then used to extract the best-fit parameters and allowed regions. 

\indent A $\chi^2$ minimization is carried out using a Markov Chain \cite{braemaud}. This minimization procedure relies on calculating the $\chi^2$ difference between successive sets of parameters and using that as a measure of whether the new point in parameter space is a ``good'' point to step to, or whether a new set of parameters needs to be drawn again. This is realized in the form of a probability of accepting a new set of parameters, $P(x_i\to x_{i+1})$, given by
\begin{equation}
P(x_i\to x_{i+1}) = min(1,e^{-(\chi^2_{i+1}-\chi^2_i)/T}),
\end{equation}
\noindent where $x_i$ and $x_{i+1}$ are two successive points in parameter space, and T is a Temperature parameter. Larger T values allow for larger $\Delta\chi^2$ jumps on the $\chi^2$ surface, and therefore by varying the T value, one can avoid local minima, as well as finely scan the parameter space. This minimization method is particularly preferred in fits with large parameter space dimensionality, as in the case of (3+2) oscillation fits, due to its higher efficiency.

\indent In extracting the various confidence level contours, we marginalize over the parameter space and report results obtained with $\Delta\chi^2$ levels corresponding to 1 degree of freedom for exclusion limits, and 2 degrees of freedom for allowed regions. 

\indent To quantify the statistical compatibility between various data sets, we use the Parameter Goodness-of-fit (PG) test introduced in \cite{Maltoni:2003cu}. In this test one quantifies how well various data sets are in agreement, by comparing the minimum $\chi^2$ obtained by a fit where all data sets have been included as constraints, $\chi^2_{min,all}$, to the sum of the $\chi^2$ minima obtained by independent fits for each experiment, i.e.,
\begin{equation}
\chi^2_{PG}=\chi^2_{min,all} - \sum_{i}\chi^2_{min,i},
\end{equation}  
\noindent where $i$ runs over experimental data sets in consideration. The PG is obtained from $\chi^2_{PG}$ based on the number of common underlying fit parameters, $ndf_{PG}$:
\begin{equation}
PG = prob(\chi^2_{PG},ndf_{PG}).
\end{equation}
\noindent For example, for testing the compatibility between KARMEN and LSND for the (3+2) CP-conserving oscillation hypothesis, we fit for both KARMEN and LSND simultaneously to extract $\chi^2_{min,K+L}$,and for KARMEN and LSND separately to extract $\chi^2_{min,K}$, and $\chi^2_{min,L}$, respectively, and obtain: 
\begin{equation}
\chi^2_{PG}(K,L) = \chi^2_{min,K+L} - (\chi^2_{min,K} + \chi^2_{min,L}).
\end{equation}
\noindent As these are appearance experiments, there are 4 common fit parameters for a CP-conserving (3+2) model ($\Delta m^2_{41}$, $\Delta m^2_{51}$, $|U_{e4}||U_{\mu4}|$, and $|U_{e5}||U_{\mu5}|$); therefore,
\begin{equation}
PG(K,L)=prob(\chi^2_{PG}(K,L),4).
\end{equation}
\indent It should be noted that $\chi^2$-probabilities and PG tests can lead to drastically different results \cite{Maltoni:2003cu}. This is often a consequence of a large data set simultaneously fitted with small data sets, where the large data set dominates the $\chi^2$ of the simultaneous fit.

\subsection{\label{sec:threea} Inclusion of the MiniBooNE Neutrino and Antineutrino Data Sets}

\indent The MiniBooNE neutrino data set (BNB-MB($\nu$)), described in Sec.~\ref{sec:mbresults}, is included in the fits in the form of two side-by-side distributions of $\nu_e$ and $\nu_{\mu}$ charged-current quasi-elastic (CCQE) events. Each distribution is a function of neutrino energy, reconstructed under the hypothesis of CCQE neutrino interaction kinematics, $E^{QE}_{\nu}$. The full 200-3000 MeV range of $\nu_e$ CCQE data is used in the fit. The observed event distributions are compared to the corresponding Monte Carlo predicted distributions, and a $\chi^2$ is calculated using a covariance matrix which includes systematic and statistical uncertainties as well as systematic correlations between the predicted $\nu_e$ and $\nu_{\mu}$ distributions.

\indent During the fit, we vary the $\nu_e$ distribution according to the sterile neutrino oscillation parameters, but keep the $\nu_{\mu}$ distribution unchanged. The $\nu_{\mu}$ distribution remains unchanged during the fit, despite the possibility of $\nu_{\mu}$ disappearance in the MiniBooNE data. In fact, MiniBooNE has released results from $\nu_{\mu}$ and $\bar{\nu}_{\mu}$ disappearance searches at $\Delta m^2\sim$ 1 eV$^2$ \cite{mbnumudis}. These results are relevant as constraints to sterile neutrino mixing parameters under consideration, but they have been purposefully omitted in this analysis, due to the fact that the $\nu_{\mu}$ and $\bar{\nu}_{\mu}$ CCQE samples used in the disappearance analysis \cite{mbnumudis} and the (different) $\nu_{\mu}$ and $\bar{\nu}_{\mu}$ CCQE samples used as constraints in the appearance analyses \cite{mblowe,mbnubar} are highly correlated samples, and these correlations have not yet become available. We assume that including MiniBooNE $\nu_{\mu}$ disappearance would have a small effect on sterile neutrino fit results, given the large overlap of the $\nu_{\mu}$ disappearance limit from MiniBooNE with the corresponding limits from CDHS and CCFR84 \cite{mbnumudis}. However, the impact of the MiniBooNE disappearance results on the fits considered in this paper will be discussed. Nevertheless, we employ this side-by-side fitting method as it takes advantage of correlations in the $\nu_{\mu}$ and $\nu_e$ predictions and in order to effectively constrain the $\nu_e$ prediction and reduce systematic uncertainties in the $\nu_{\mu}\to\nu_e$ search. 

\indent The fit method follows the details described in \cite{mblowe}, except that it uses a different definition for the covariance matrix used in the $\chi^2$ calculation. Ref.~\cite{mblowe} involves an iterative fit method where the $\chi^2$ calculation for each point on the parameter space being probed uses the covariance matrix calculated according to the {\it best-fit} signal prediction. Instead, in the MiniBooNE fits presented here, the $\chi^2$ surface is estimated using the covariance matrix calculated according to the signal prediction at {\it each point} of the parameter space under consideration. The two fit methods yield similar results, although, by definition, the iterative method of \cite{mblowe} results in a relative shift of the allowed region to the left, i.e.~towards smaller oscillation amplitudes.

\indent The MiniBooNE antineutrino data set (BNB-MB($\bar{\nu}$)), described in Sec.~\ref{sec:mbresults}, is included in the fits in the same way as the BNB-MB($\nu$) data set, in the form of two side-by-side $E^{QE}_{\nu}$ distributions of $\bar{\nu}_e$ and $\bar{\nu}_{\mu}$ CCQE events. In this case, the disappearance limit obtained using the MiniBooNE $\bar{\nu}_{\mu}$ CCQE sample provides substantial coverage of so-far unexplored sterile neutrino mass and mixing parameter space \cite{mbnumudis}. Even though we do not explicitly fit the MiniBooNE $\bar{\nu}_{\mu}$ CCQE distribution for disappearance, we comment on the effect of the limit from \cite{mbnumudis} in Sec.~\ref{sec:four}, and justify that excluding the MiniBooNE $\bar{\nu}_{\mu}$ disappearance information from the fits does not substantially affect the parameter space of interest. The full 200-3000 MeV range of $\bar{\nu}_e$ CCQE data is used in the fit. The BNB-MB($\bar{\nu}$) data fit method also follows the details described in \cite{mbnubar}, except that it uses the definition for the covariance matrix described above.

\indent In fits where both neutrino and antineutrino data are included, it has been assumed that the two data sets are fully uncorrelated. In reality, the two data sets have large systematic correlations. However, neglecting the effects of these correlations is a reasonable approximation, given that the antineutrino data set is statistics limited. 

\subsection{\label{sec:threeb} Inclusion of the NuMI-beam Data Set}

\indent The NUMI-MB data set \cite{numi}, described in Sec.~\ref{sec:mbresults}, is included in the fits in the form of a distribution of $\nu_e$ CCQE events as a function of reconstructed neutrino energy, $E^{QE}_{\nu}$. The predicted $\nu_e$ distribution is obtained by adding to the expected $\nu_e$ CCQE background any contribution from $\nu_{\mu}$ to $\nu_e$ oscillations. The contribution is estimated as follows: First, a fully (100\%) oscillated NUMI-MB $\nu_{\mu}\to\nu_e$ sample is obtained by reweighting the BNB-MB fully oscillated $\nu_{\mu}\to\nu_e$ Monte Carlo predicted sample according to the ratio of the NuMI-beam flux from \cite{numi} to the BNB-MB flux \cite{mbflux}, as a function true neutrino energy. As the oscillation parameters vary during the fit, a signal prediction is calculated by rescaling the number of events in this fully oscillated sample by the corresponding oscillation probability, according to the true neutrino energy and distance travelled, from production to detection, of each event. We assume a constant $L$ of 700 meters. The prediction is compared to the observed $\nu_e$ CCQE events as a function of 10 bins of $E^{QE}_{\nu}$. The background and signal prediction are assumed to have the same fractional systematic uncertainties, and a statistical uncertainty is calculated for each point in the parameter space according to the signal prediction of each point under consideration. The data and background central value and systematic uncertainty per $E^{QE}_{\nu}$ bin have been estimated from \cite{numi}. Unlike the systematic uncertainties of the BNB-MB $\nu_{e}$ and $\bar{\nu}_e$ CCQE data sets, the NUMI-MB $\nu_e$ CCQE systematics have not been constrained using information from the $\nu_{\mu}$ CCQE spectrum from the NuMI beamline. Furthermore, we have not considered potential systematic correlations among the $\nu_e$ CCQE bins as a function of $E^{QE}_{\nu}$.


\section{\label{sec:four}(3+1) AND (3+2) MODELS AFTER THE NEW MINIBOONE $\nu$, $\bar{\nu}$, AND NUMI RESULTS}

\subsection{(3+1) FIT RESULTS}

\indent In this section, the new MiniBooNE results are examined under a (3+1) oscillation hypothesis and compared to LSND and other null SBL experiments. The new data sets are studied first within the context of appearance-only experiments, and subsequently in fits involving both appearance and disappearance data. Fits to only antineutrino and only neutrino SBL experiments are also explored.

\begingroup
\squeezetable
\begin{table*}[tb] 
\begin{ruledtabular} 
\begin{tabular}{l|ccccccccccc} 
Fit              &               &                     &         &         & Data Sets &       &       &       &      &      &    \\
                 & BNB-MB($\nu$) & BNB-MB($\bar{\nu}$) & LSND    & NUMI-MB & KARMEN  & NOMAD   & CHOOZ   & Bugey   & CCFR    & CDHS    & atm\\\hline\hline
APP              &  $\surd$      &       $\surd$       & $\surd$ & $\surd$ & $\surd$ & $\surd$ &         &         &         &         &         \\
DIS              &               &                     &         &         &         &         & $\surd$ & $\surd$ & $\surd$ & $\surd$ & $\surd$ \\
$\nu$            &  $\surd$      &                     &         & $\surd$ &         & $\surd$ &         &         & $\surd$ & $\surd$ & $\surd$ \\
$\nu$ APP        &  $\surd$      &                     &         & $\surd$ &         & $\surd$ &         &         &         &         &         \\
$\bar{\nu}$      &               &       $\surd$       & $\surd$ &         & $\surd$           & $\surd$ & $\surd$ &         &         &         \\
$\bar{\nu}$ APP  &               &       $\surd$       & $\surd$ &         & $\surd$ &         &         &         &         &         &         \\
signal           &  $\surd$      &       $\surd$       & $\surd$ &         &         &         &         &         &         &         &         \\
signal APP       &  $\surd$      &       $\surd$       & $\surd$ &         &         &         &         &         &         &         &         \\
null             &               &                     &         & $\surd$ & $\surd$ & $\surd$ & $\surd$ & $\surd$ & $\surd$ & $\surd$ & $\surd$ \\
null APP         &               &                     &         & $\surd$ & $\surd$ & $\surd$ &         &         &         &         &         \\  
\end{tabular} 
\end{ruledtabular} 
\caption{\label{classification} Short-baseline oscillation fits considered in this paper.}
\end{table*} 
\endgroup

\begin{figure*}[htb] 
\includegraphics[ width=17cm]{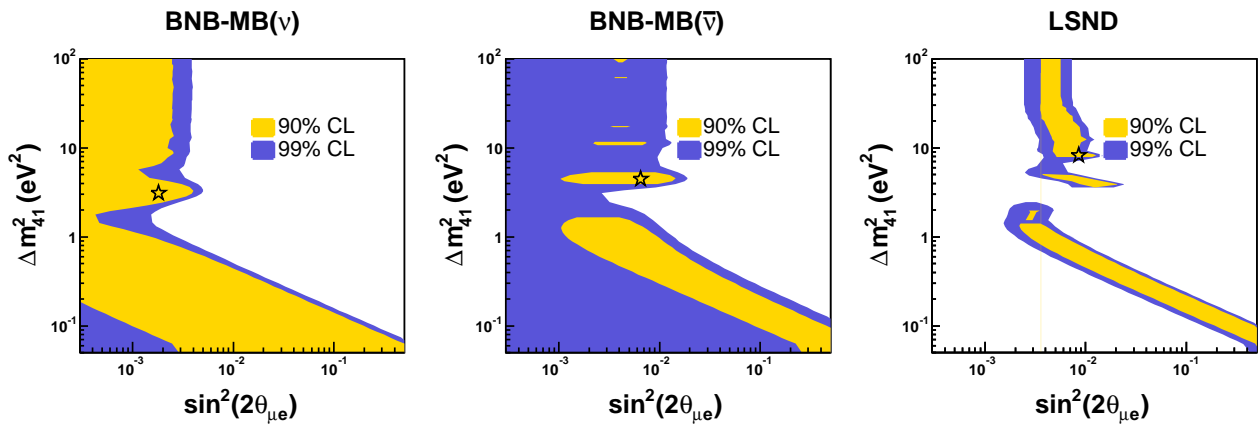} 
\caption{\label{fig1} Allowed regions (filled areas) at 90\% and 99\% CL from BNB-MB($\nu$)-only, BNB-MB($\bar{\nu}$)-only, and LSND-only (3+1) fits. These fits are, by construction, CP-conserving. The stars indicate the three respective best-fit points. All three data sets show closed contours at 90\% CL. See text for more details. }
\end{figure*}

\begingroup
\squeezetable
\begin{table*}[tb] 
\begin{ruledtabular} 
\begin{tabular}{lccclll} 
Data Set & $\chi^2\ (dof)$ & $\chi^2$-probability & $\Delta m^2_{41}$ &  $\sin^2 2\theta_{\mu e}$ & $\sin^2 2\theta_{\mu\mu}$ & $\sin^2 2\theta_{ee}$ \\ \hline \hline
{\bf Appearance-only fits:} & & & & & & \\
LSND                            & 3.4 (3)         & 34\% & 8.19 & 0.0085 & - & - \\
BNB-MB($\nu$)                   & 17.5 (16)       & 35\% & 3.12 & 0.0018 & - & -  \\
BNB-MB($\bar{\nu}$)             & 17.6 (16)       & 35\% & 4.46 & 0.0065 & - & -  \\
NUMI-MB                         & 2.0 (8)         & 98\% & 6.97 & 0.020  & - & -  \\
KARMEN                          & 6.0 (7)         & 54\% & 6.81 & 0.00096 & - & - \\
NOMAD                           & 33.3 (28)       & 31\% & 53.3 & 0.00012 & - & - \\
\hline
signal APP                      & 50.3 (39)       & 11\% & 0.045 & 0.98 & - & - \\
signal APP$^{*}$                & 50.4 (39)       & 10\% & 0.15 & 0.090  & - & - \\
null APP                        & 46.6 (47)       & 49\% & 0.040 & 1.00 & - & - \\
APP                             & 97.1 (88)       & 24\% & 0.045 & 1.00  & - & -  \\
APP$^{*}$                       & 97.2 (88)       & 24\% & 0.15 & 0.090  & - & - \\
\hline
LSND + MB-BNB($\bar{\nu}$)      & 22.3 (21)       & 38\% & 4.55 & 0.0074 & - & - \\
LSND + MB-BNB($\bar{\nu}$)$^{*}$ & 22.3 (21)      & 38\% & 4.55 & 0.0074 & - & -  \\
LSND + MB-BNB($\bar{\nu}$) + KARMEN   & 33.6 (30) & 29\% & 0.57 & 0.0097 & - & - \\
BNB-MB($\nu$) + NUMI-MB + NOMAD       & 57.8 (56) & 40\% & 0.033& 1.00  & - & - \\
\hline
{\bf Appearance and disappearance fits:} & & & & & & \\
all SBL$^{*}$                   & 197.4 (196)     & 46\% & 0.92 & 0.0025  & 0.13 & 0.073 \\
$\nu$                           & 90.5 (90)       & 47\% & 0.19 & 0.031   & 0.031 &  0.034 \\
$\bar{\nu}$                     & 87.9 (103)      & 86\% & 0.91 & 0.0043 & 0.35 & 0.043 \\
\end{tabular} 
\end{ruledtabular} 
\caption{\label{tab:3plus1fits} Comparison of best-fit values for mass splittings and mixing angles obtained from   
 (3+1) fits to appearance data sets and appearance+disappearance data sets. Mass splittings are shown in eV$^2$. The minimum $\chi^2$ from each fit, as well as the $\chi^2$-probability are also given. The signal appearance (APP) data sets include BNB-MB($\nu$), BNB-MB($\bar{\nu}$) and LSND. The null APP data sets include KARMEN, NOMAD and NUMI-MB; the maximal best-fit $\sin^2 2\theta_{\mu e}$ in this case is inconsequential, as it corresponds to a best-fit $\Delta m^2$ region of very poor sensitivity. See text for more details.\\
$^*${\it In these fits, the electron and muon content of the sterile neutrino mass eigenstate have been explicitly constrained to $<$0.3.}} 
\end{table*} 
\endgroup

\begin{figure*}[htb] 
\vspace{0.5cm}
\includegraphics[ width=6cm, angle=-90, trim=50 60 0 50]{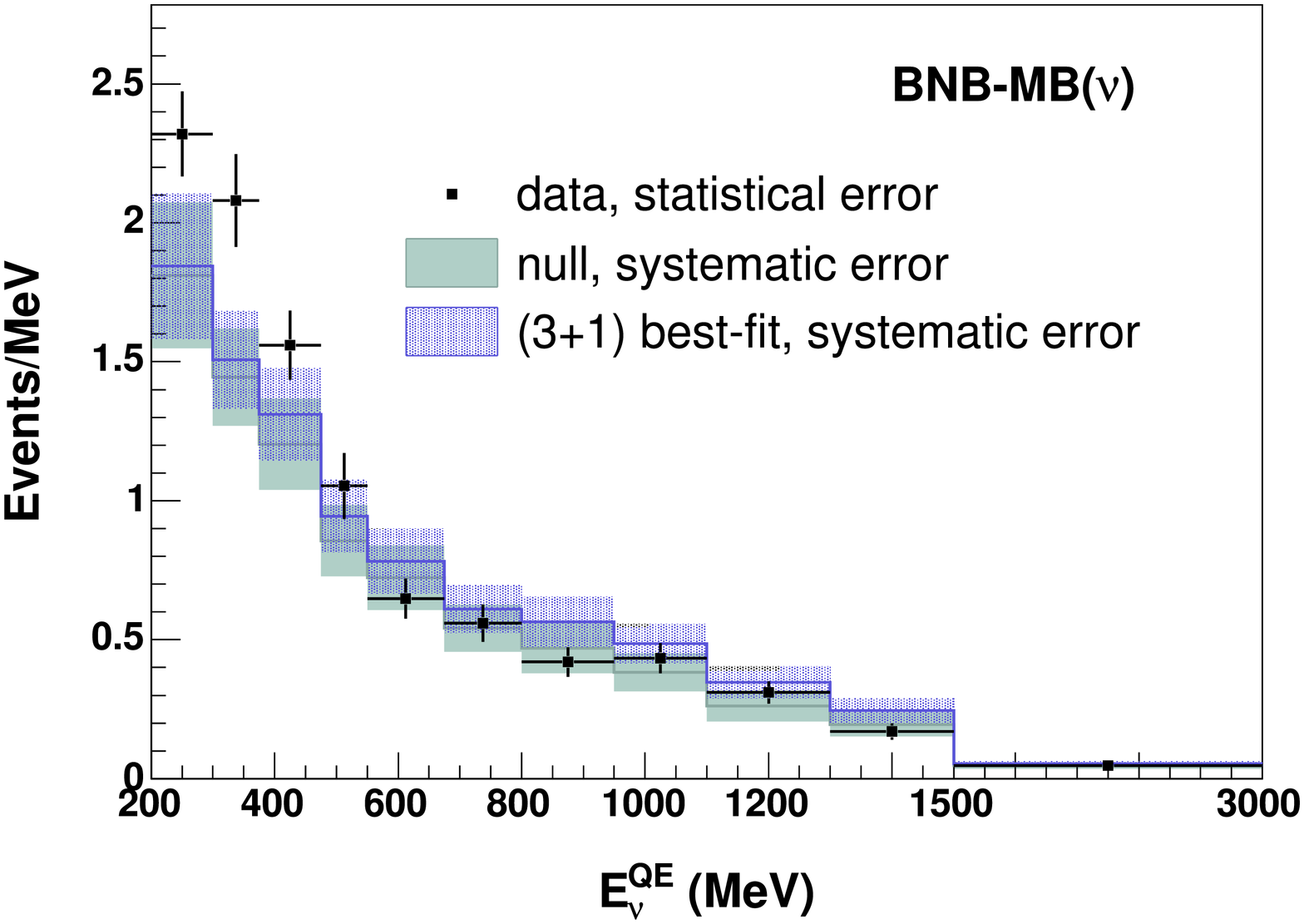} 
\hspace{1cm}
\includegraphics[ width=6cm, angle=-90, trim=50 60 0 50]{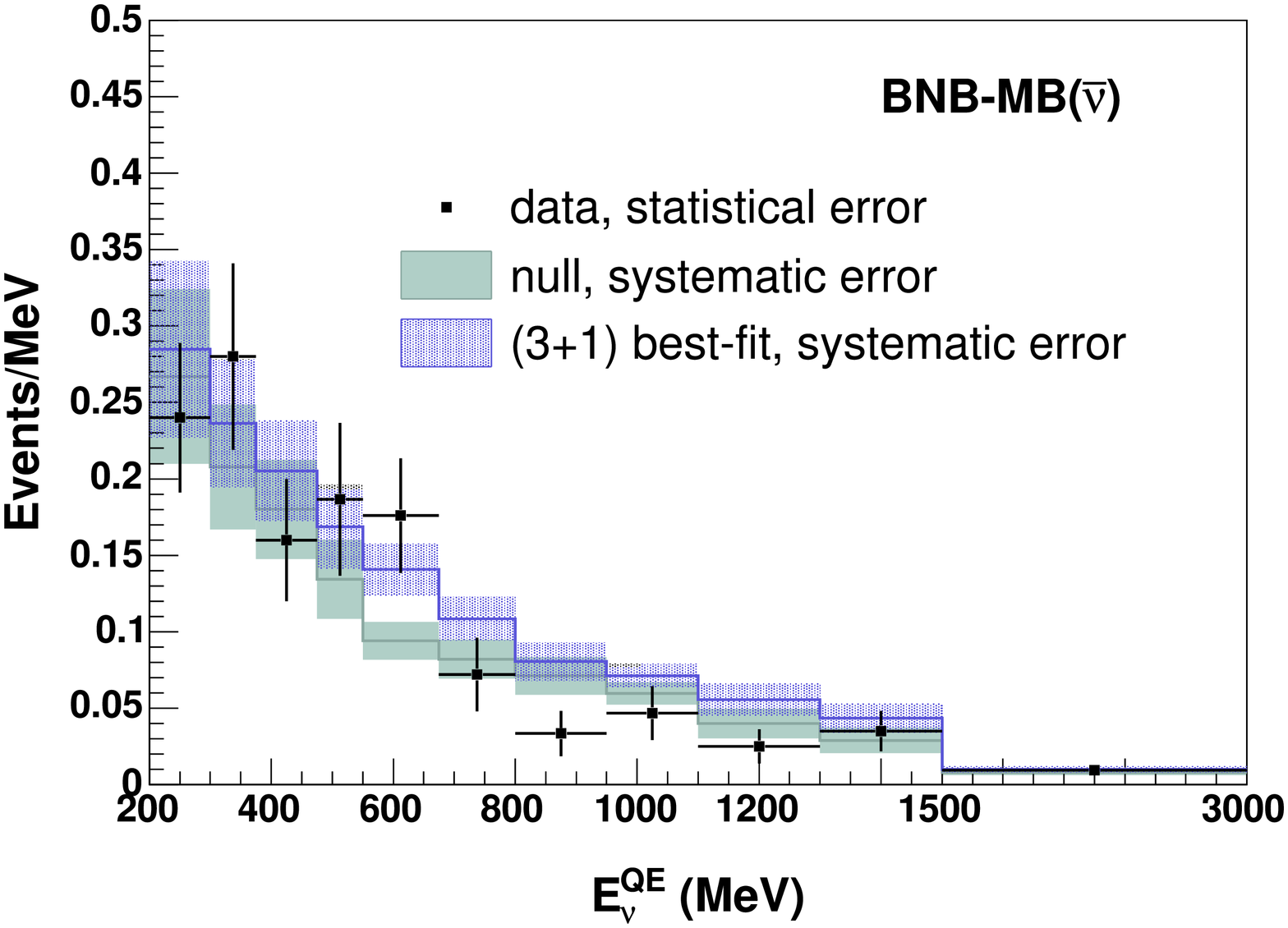} 
\caption{\label{fig1pred} Left: Null and (3+1) best-fit predicted event distributions ($\Delta m_{41}^2,\sin^2(2\theta_{\mu e})$) $=$ (3.12, 0.0018) for BNB-MB($\nu$). Right: Null and (3+1) best-fit predicted event distributions ($\Delta m_{41}^2,\sin^2(2\theta_{\mu e})$) $=$ (4.46, 0.0065) for BNB-MB($\bar{\nu}$). The event distributions are shown as functions of reconstructed neutrino energy, $E^{QE}_{\nu}$. The data are shown in black points with statistical uncertainty. The null (no-oscillation) prediction is shown by the light gray histogram with (solid) systematic error band. The best-fit prediction (signal and background) is shown by the blue (dark gray) histogram with (shaded) systematic error band.}
\end{figure*}

\indent Table \ref{classification} provides a reference for all the different data set combinations explored in fits in this paper.

\subsubsection{Studies with appearance-only experiments}

\indent Figure \ref{fig1} shows the allowed regions obtained by independent fits to each of the following three data sets: BNB-MB($\nu$), BNB-MB($\bar{\nu}$), and LSND. The regions are estimated using a 2-dimensional global scan of the (3+1) parameter space ($\sin^2 2\theta_{\mu e}$,$\Delta m^2_{41}$). Each contour is drawn by applying a flat $\Delta \chi^2=\chi^2(\sin^2 2\theta_{\mu e},\Delta m_{41}^2)-\chi^2_{min}$ cut over the $\chi^2$ surface, with respect to the global $\chi^2$ minimum returned by the fit. The figure shows that, similarly to LSND, both BNB-MB data sets yield contours which exclude the no-oscillations (null) hypothesis at 90\% CL. The null $\chi^2$'s correspond to 22.2 and 24.5 for BNB-MB($\nu$) and BNB-MB($\bar{\nu}$), respectively. The closed contours reflect a contradiction to the oscillation results published by the MiniBooNE collaboration; this is a consequence of the different $\chi^2$ definition involved in the fit method used here, as pointed out in Sec.~\ref{sec:threea}. All three data sets, BNB-MB($\nu$), BNB-MB($\bar{\nu}$), and LSND, yield similar best-fit parameters, indicated by the stars on the three graphs, of $\Delta m_{41}^2$ of order a few eV$^2$ and $\sin^2 2\theta_{\mu e}$ of order 10$^{-2}-$10$^{-3}$. The minimum $\chi^2$ and best-fit parameters returned by each experiment are summarized in Table \ref{tab:3plus1fits}. 

\indent In light of the above BNB-MB results and the already established LSND anomaly, we find it interesting to study sterile neutrino oscillations with the LSND, BNB-MB($\nu$), and BNB-MB($\bar{\nu}$) data sets assumed to be (positive) ``signal'' experiments under both the (3+1) and the (3+2) models. This classification is based on the fact that all three data sets exclude the null result at 90\% confidence level under a (3+1) oscillation hypothesis. Figure \ref{fig1pred} shows the BNB-MB($\nu$) and BNB-MB($\bar{\nu}$) event distributions for both the null and the best-fit (3+1) oscillation hypothesis for each data set. In the case of the BNB-MB($\nu$) data set, even though the best-fit hypothesis provides a better description of the event spectrum at 90\% CL ($\Delta\chi^2=\chi^2_{null}-\chi^2_{best-fit}=$4.7, for 2 fit parameters), it fails to fully explain the low energy excess. Therefore, the (3+1) oscillation hypothesis alone seems inadequate as an explanation for the low energy excess, as also reported by the MiniBooNE collaboration \cite{mbnu,mblowe}. In the case of the BNB-MB($\bar{\nu}$) data set, the best-fit hypothesis provides a better description of the data in the 500-700 MeV range. However, the statistical uncertainties are too large to allow for a strong conclusion.

\begingroup
\squeezetable
\begin{table*}[tb] 
\begin{ruledtabular} 
\begin{tabular}{l|c|lcc} 
Data Set & $\chi^2$-probability (\%) & \ \ \ \ \ \ \ \ \  PG (\%) & & \\ \hline \hline
APP                                  & 24 & PG( BNB-MB($\nu$),BNB-MB($\bar{\nu}$),LSND,NUMI-MB,KARMEN,NOMAD ) & = Prob( 17.3,10 ) = & 6.8 \\
signal APP                           & 11 & PG( BNB-MB($\nu$),BNB-MB($\bar{\nu}$),LSND )                      & = Prob( 11.9,4 ) = & 1.8 \\
LSND + MB-BNB($\bar{\nu}$)           & 38 & PG( BNB-MB($\bar{\nu}$),LSND )                                    & = Prob( 1.4,2 ) = & 49.0 \\
$\bar{\nu}$ APP                      & 29 & PG( BNB-MB($\bar{\nu}$),LSND,KARMEN )                             & = Prob( 6.7,4 ) = & 15.3 \\
$\nu$ APP                            & 40 & PG( BNB-MB($\nu$),NUMI-MB,NOMAD )                                 & = Prob( 4.9,4 ) = & 29.8 \\
\hline
all SBL$^{*}$                        & 46 & PG( BNB-MB($\nu$),BNB-MB($\bar{\nu}$),NUMI-MB,LSND,KARMEN, &  &   \\
                                     &    & \ \ \ \ \ \ \ NOMAD,Bugey,CHOOZ,CCFR84,CDHS,ATM )   & = Prob( 42.0,18 ) = & 0.11 \\
                                     &    & PG( APP,DIS )                                       & = Prob( 14.8,2 ) = & 0.06 \\
                                     &    & PG( $\nu$,$\bar{\nu}$ )                             & = Prob( 18.8,3 ) = & 0.03 \\
\hline
$\nu$                            & 47 & PG( BNB-MB($\nu$),NUMI-MB,NOMAD,CCFR84,CDHS,ATM )   & = Prob( 14.7,8 ) = & 6.5 \\
$\bar{\nu}$                      & 86 & PG( BNB-MB($\bar{\nu}$),LSND,KARMEN,Bugey,CHOOZ )   & = Prob( 8.43,7 ) = & 29.9 \\
\end{tabular} 
\end{ruledtabular} 
\caption{\label{tab:3plus1fitsPG}  Summary of $\chi^2$-probabilities for (3+1) fits with different combinations of SBL data sets, and PG results testing compatibility among different data sets. See text for more details.\\
$^*${\it In these fits, the electron and muon content of the sterile neutrino mass eigenstate have been explicitly constrained to $<$0.3.}} 
\end{table*} 
\endgroup

\begin{figure*}[htb] 
\vspace{1.0cm}
 \includegraphics[width=5.5cm, trim=55 55 55 55]{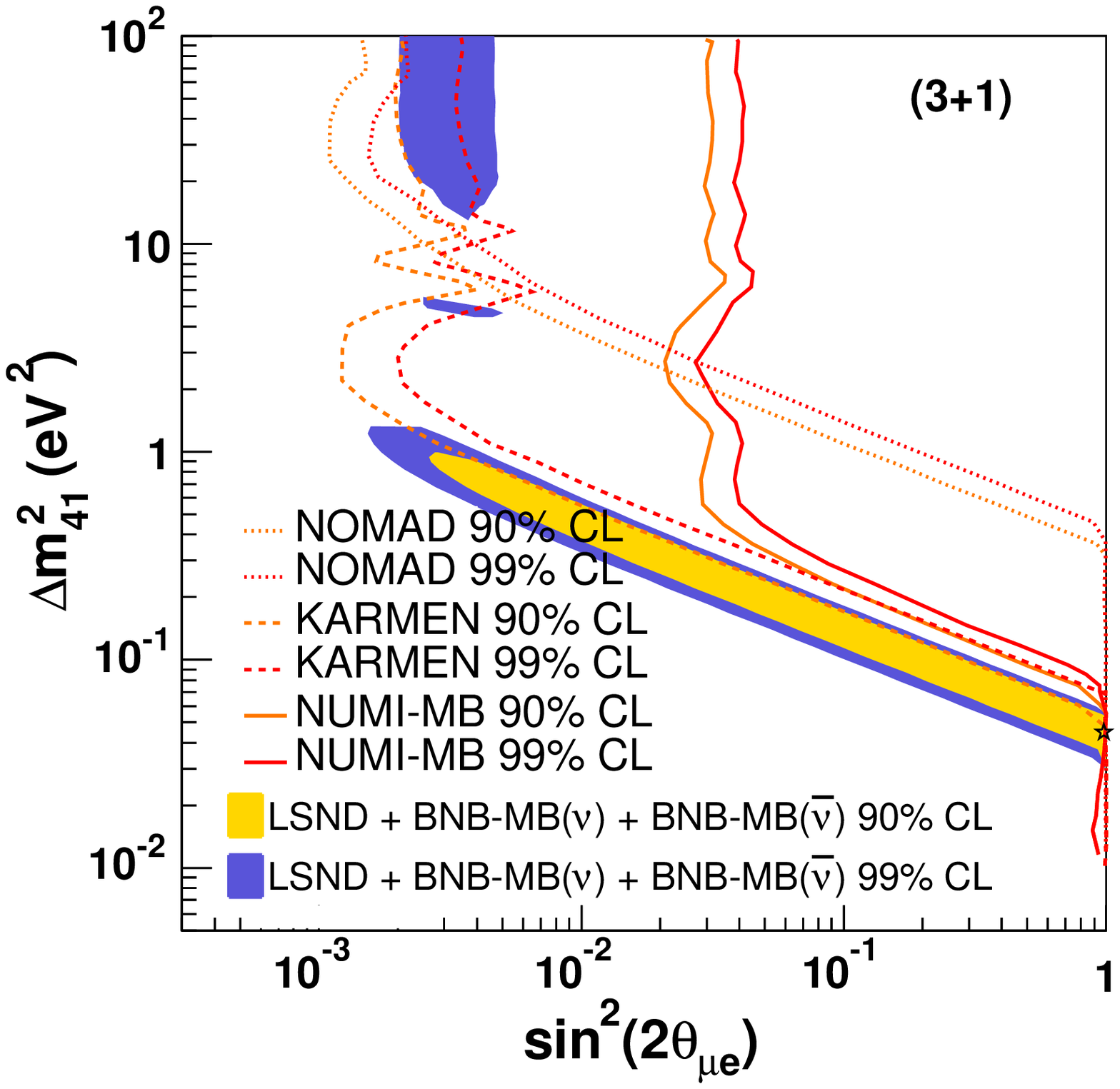} 
\hspace{0.3cm}
 \includegraphics[width=5.5cm, trim=55 55 55 55]{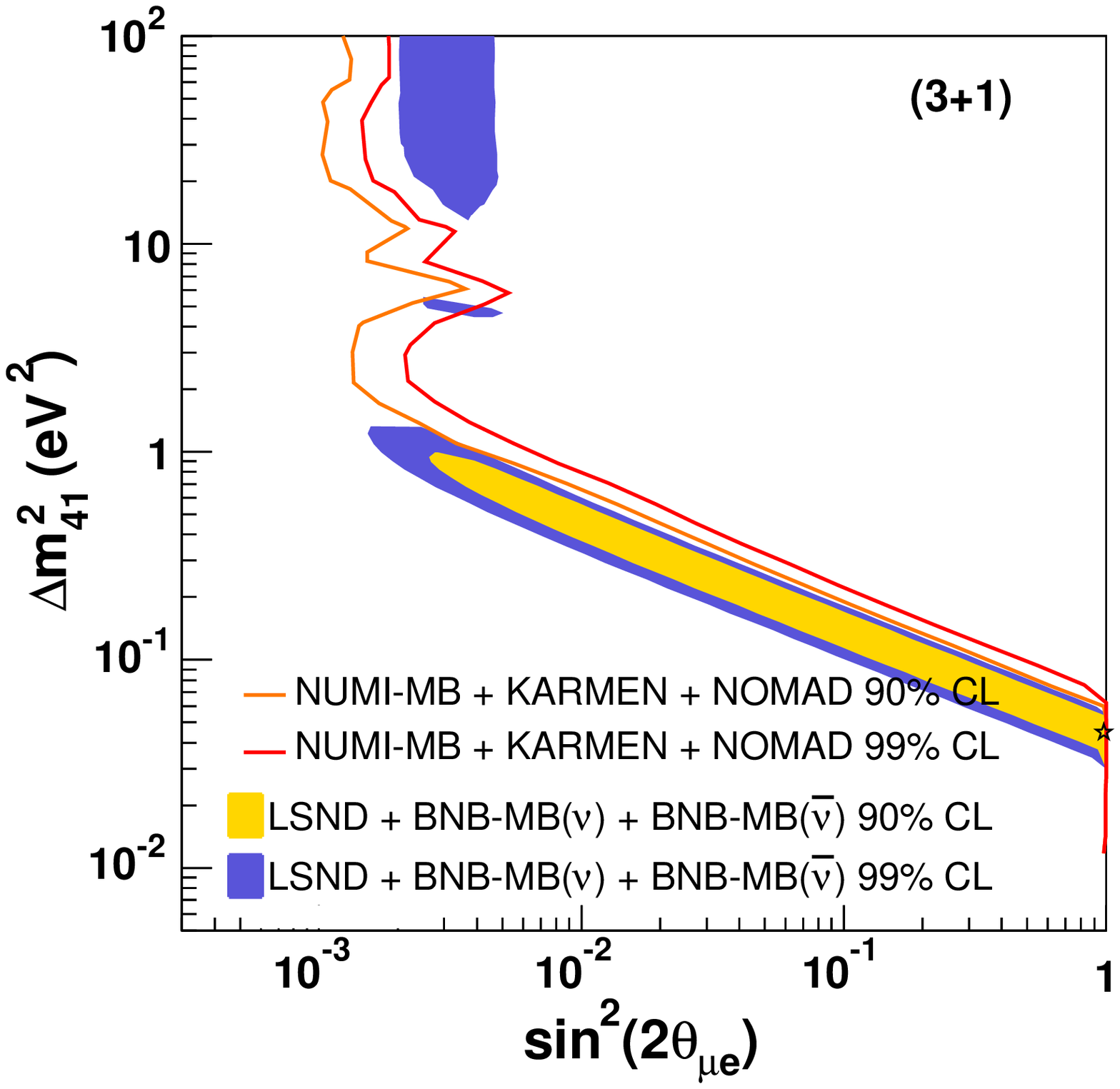} 
\hspace{0.3cm}
 \includegraphics[width=5.5cm, trim=55 55 55 55]{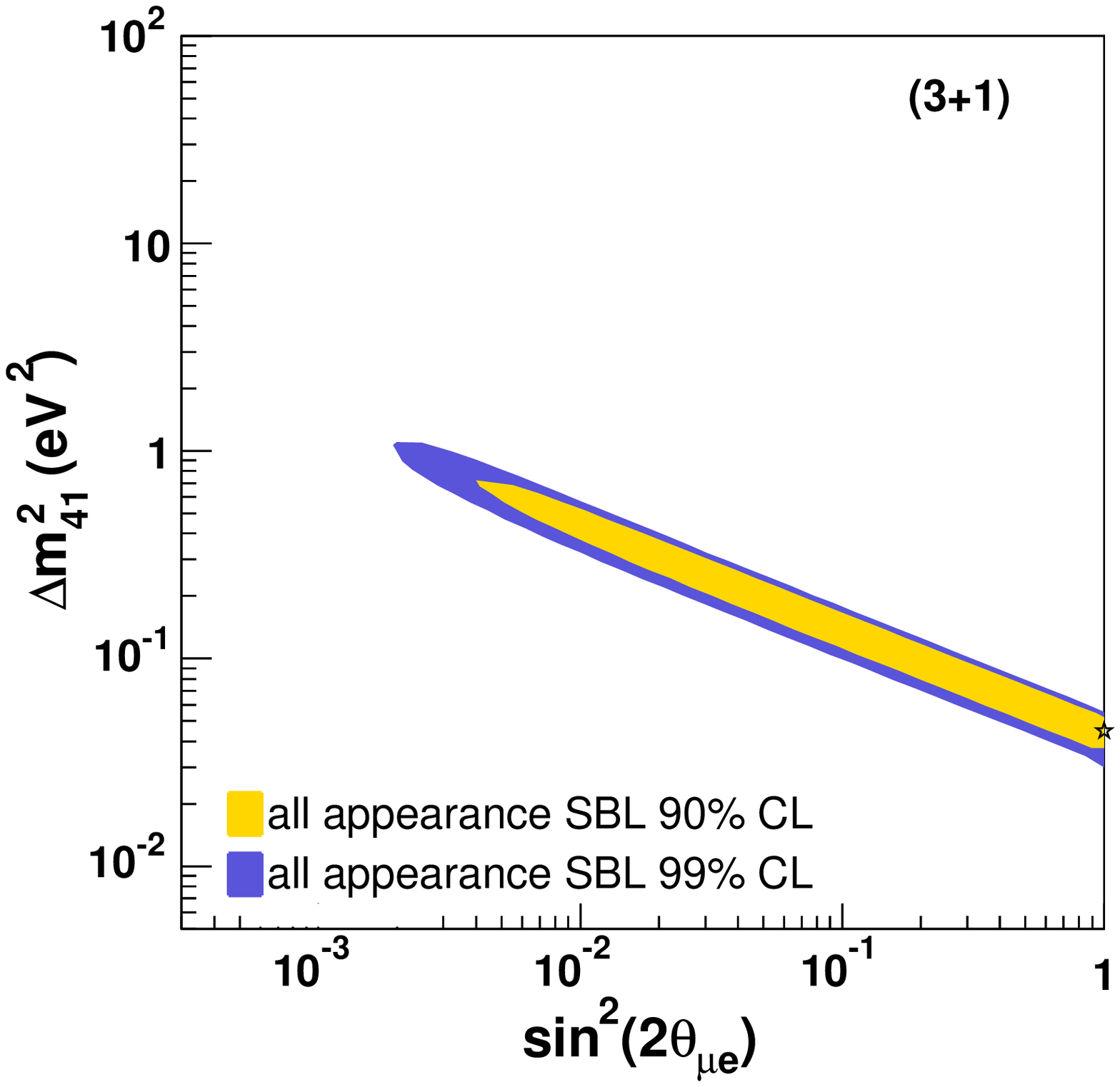} 
\vspace{1.0cm}
 \caption{\label{fig2a} Left: Allowed 90\% and 99\% CL regions (light and dark filled areas, respectively) from a combined analysis of BNB-MB($\nu$), BNB-MB($\bar{\nu}$) and LSND data sets, and 90\% and 99\% exclusion limits (light and dark curves, respectively) from each of the null appearance experiments, NUMI-MB (solid curves), KARMEN (dashed curves) and NOMAD (dotted curves). Middle: The same allowed region with overlayed 90\% and 99\% exclusion limits from a combined analysis of all null appearance experiments. Right: Allowed region obtained by a combined analysis of all appearance data sets, signal and null. See text for more details.}
\end{figure*}

\begin{figure*}[htbp]
\vspace{1.5cm}
 \includegraphics[width=5.5cm, trim=55 55 55 55]{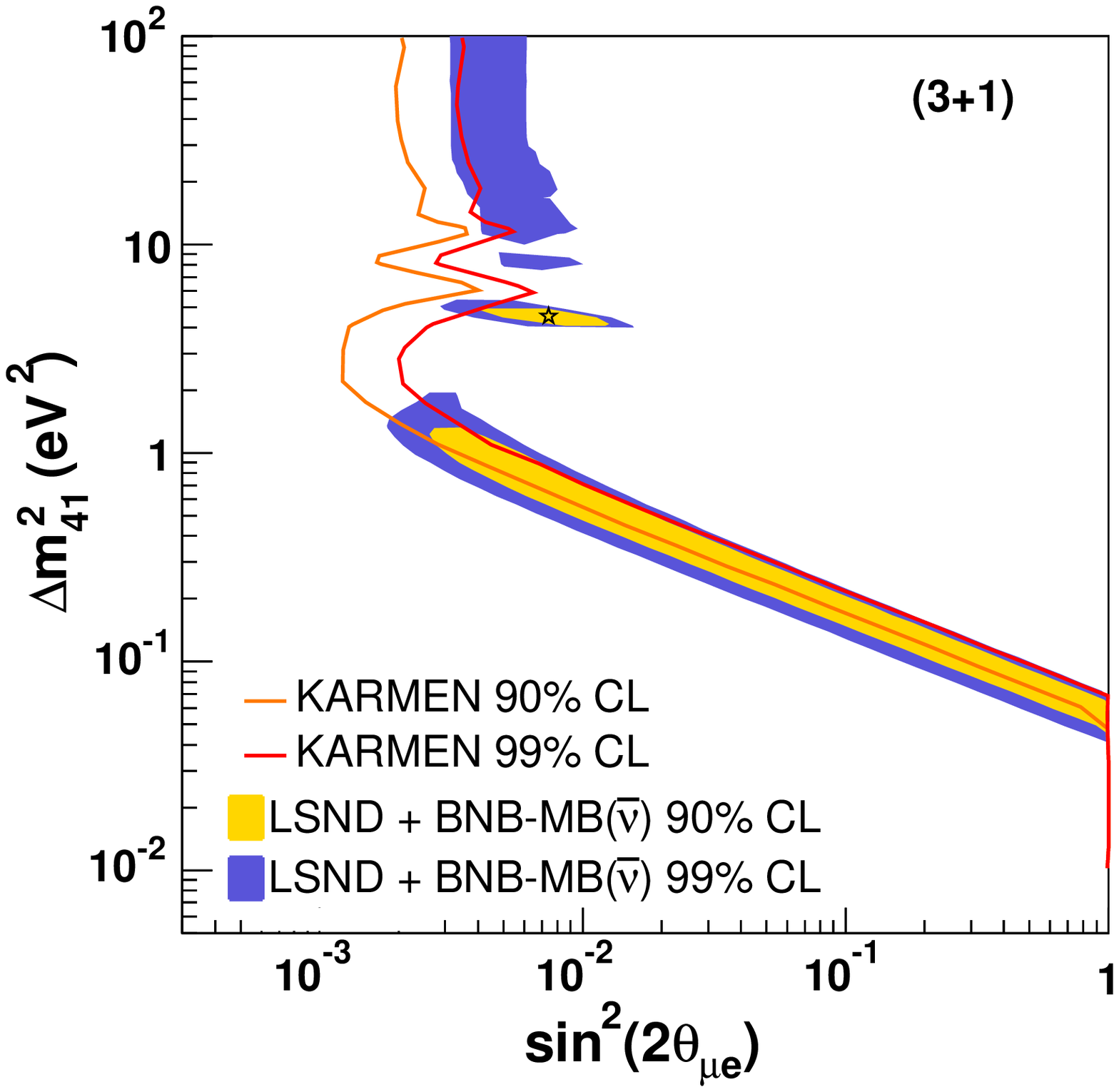} 
\hspace{1cm}
 \includegraphics[width=5.5cm, trim=55 55 55 55]{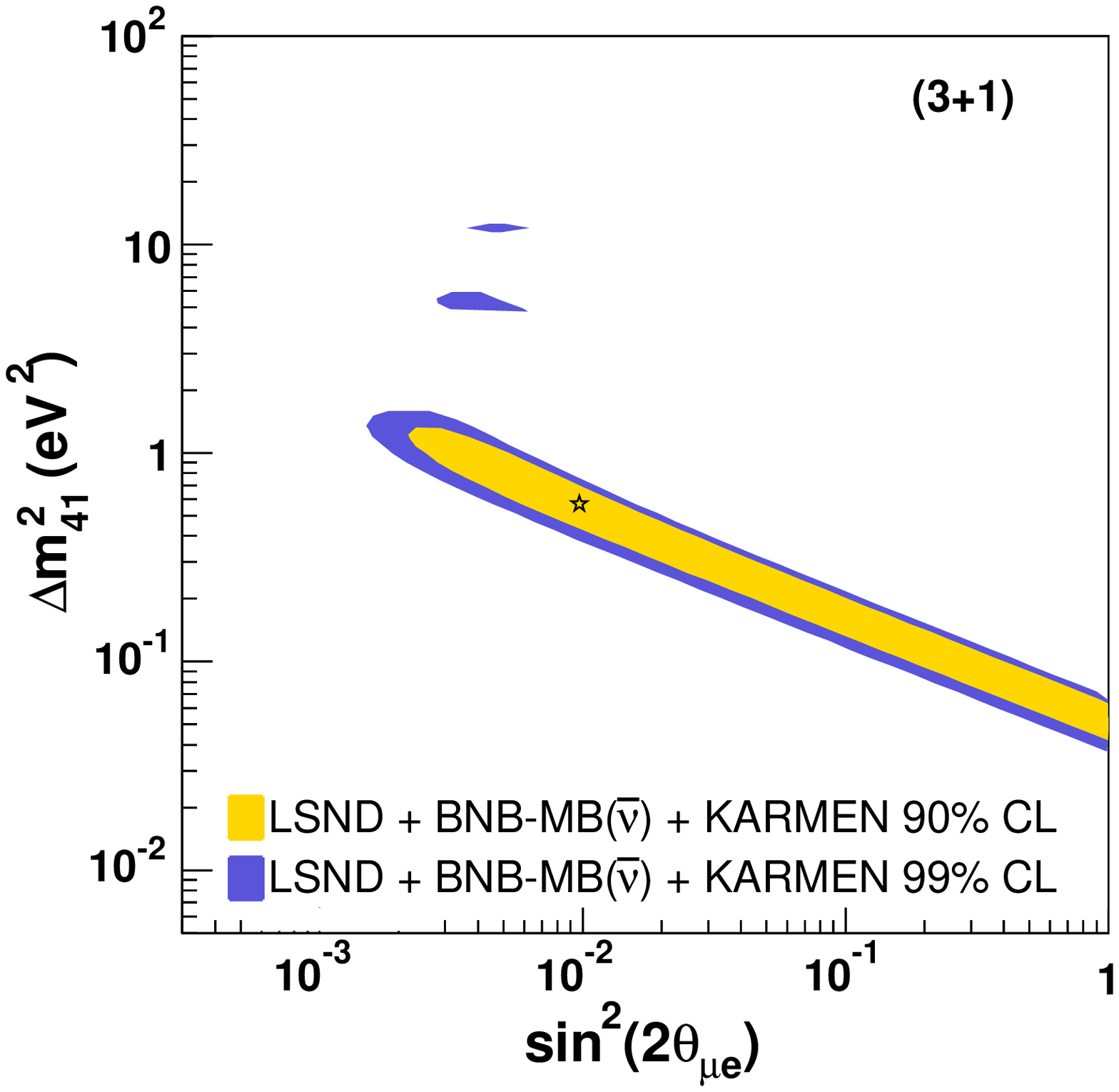} 
\vspace{1.0cm}
 \caption{\label{fig2b} Left: The allowed 90\% and 99\% CL regions (light and dark filled areas, respectively) from a combined analysis of BNB-MB($\bar{\nu}$) and LSND data sets, and 90\% and 99\% exclusion limits (light and dark curves, respectively) from KARMEN. A comparison of only these three experiments is interesting, as these three experiments have searched for antineutrino oscillations at short baselines. Right: The allowed regions obtained from a combined analysis of all three experiments (BNB-MB($\bar{\nu}$), LSND, and KARMEN). See text for more details.}
\end{figure*}

\indent The allowed regions obtained by a joint analysis of BNB-MB($\nu$) + BNB-MB($\bar{\nu}$) + LSND, as well as a joint analysis of BNB-MB($\bar{\nu}$) + LSND are shown on the left panels of Figs.~\ref{fig2a} and \ref{fig2b}, respectively. In the case of the combined fit of all three data sets (Fig.~\ref{fig2a}), due to the difference in preferred mixing amplitudes mentioned in the previous paragraph, the best-fit point ends up shifting from an intermediate $\Delta m^{2}$ and small mixing amplitude to a smaller $\Delta m^2$ and maximal mixing amplitude of 0.98. Obviously a maximal mixing amplitude is unphysical in the case of sterile neutrino oscillations. If the fits are repeated with the electron and muon content of the sterile mass eigenstate limited to values less than 0.3 \footnote{This constraint is imposed under the assumption that mixing with sterile neutrinos should be small.}, the returned $\chi^2$-probabilities are 10\% and 38\%, for BNB-MB($\nu$) + BNB-MB($\bar{\nu}$) + LSND and BNB-MB($\bar{\nu}$) + LSND, respectively; the reduction in $\sin^2 2\theta_{\mu e}$ space has essentially no effect on these results. The best-fit parameters from these fits are also given in Table \ref{tab:3plus1fits}.

\indent Perhaps a more interesting observation regarding Fig.~\ref{fig1} is the striking similarity of BNB-MB($\bar{\nu}$) and LSND 90\% CL allowed regions and best-fit oscillation parameters, keeping in mind that both data sets describe antineutrino oscillations. It should be noted that in a (3+1) oscillation scenario, under the assumption of CPT invariance, there can be no difference between neutrino and antineutrino oscillation probabilities. However, a PG test, as described in Sec.~\ref{sec:three0}, suggests a significantly higher compatibility (49\%) between BNB-MB($\bar{\nu}$) and LSND, rather than for all three signal experiments (BNB-MB($\nu$), BNB-MB($\bar{\nu}$) and LSND) combined (1.8\%). This is also supported by the $\chi^2$-probabilities returned by the fits: 11\% in the case of the BNB-MB($\nu$) + BNB-MB($\bar{\nu}$) + LSND fit, and 38\% in the case of the BNB-MB($\bar{\nu}$) + LSND fit. This incompatibility is due to the fact that the BNB-MB($\nu$) data set prefers a mixing amplitude $\sim$3 times smaller than the amplitude preferred by LSND or BNB-MB($\bar{\nu}$), and excludes the LSND and BNB-MB($\bar{\nu}$) best-fits at 99\% CL. Table \ref{tab:3plus1fitsPG} provides a summary of the above $\chi^2$-probabilities and PG test results.

\begin{figure}[htbp]
\vspace{0.4cm}
\includegraphics[ width=6cm, angle=-90, trim=50 60 0 50 ]{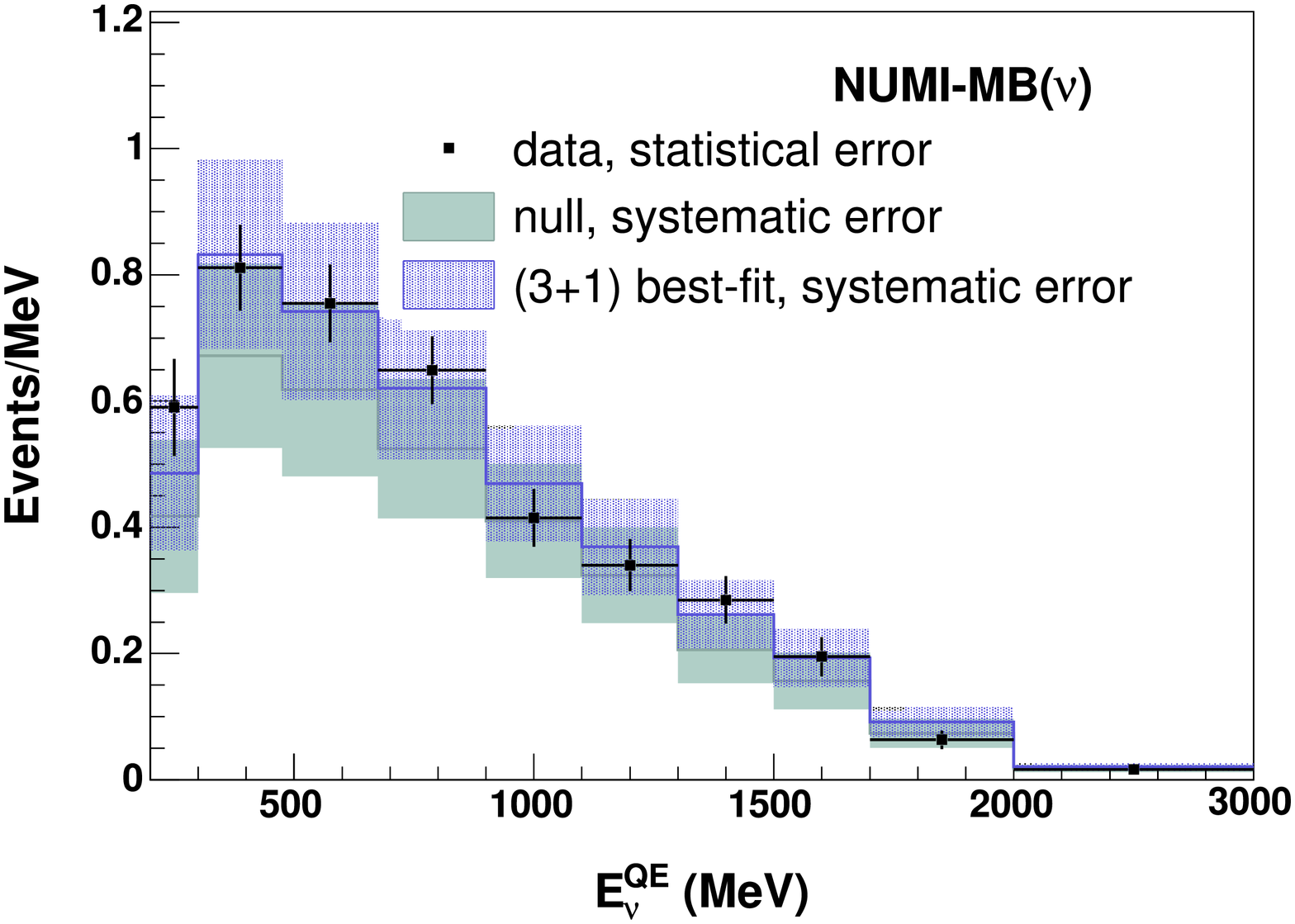} 
\caption{\label{fig2c} Null and (3+1) best-fit predicted event distributions ($\Delta m_{41}^2,\sin^2(2\theta_{\mu e})$) $=$ (7.36, 0.019) for NUMI-MB. The data are shown in black points with statistical uncertainty. The null (no-oscillation) prediction is shown by the light gray histogram with (solid) systematic error band. The best-fit prediction (signal and background) is shown by the blue (dark gray) histogram with (shaded) systematic error band.}
\end{figure}

\indent Figures \ref{fig2a} and \ref{fig2b} also illustrate the limits from various combinations of the remaining three (null) SBL appearance experiments, KARMEN, NOMAD, and NUMI-MB, under a (3+1) oscillation scenario, overlaid on the allowed regions described above. 

\indent The 90\% and 99\% CL limits obtained by each of the null appearance experiments are shown on the left panel of Fig.~\ref{fig2a}. These limits correspond to the upper $\sin^2 2\theta_{\mu e}$ values allowed at each $\Delta m_{41}^2$, estimated using a one-sided raster scan of the parameter space. It is interesting to point out that, despite the indication of a slight excess (1.2$\sigma$) of observed $\nu_e$-like events for neutrino energies below 900 MeV found in the NuMI analysis \cite{numi}, the currently assumed NUMI-MB systematic and statistical uncertainties are quite large, resulting in a limit that is much weaker relative to the limits of KARMEN and NOMAD. In fact, due to this excess and the large systematic uncertainties, the NUMI-MB data set provides a very good fit to (3+1) models, with a $\chi^2$-probability of 98\%. The event distributions for the null and best-fit (3+1) oscillation hypothesis for the NUMI-MB data set are shown in Fig.~\ref{fig2c}. The observed distribution fits nicely to an oscillation signal at ($\Delta m_{41}^2,\sin^2(2\theta_{\mu e})$) $=$ (7.36, 0.019). Such large signal, however, would be in disagreement with the BNB-MB($\nu$) results. Additional data and reduced systematic uncertainties in the NUMI-MB analysis are necessary for higher sensitivity and more conclusive results. This is currently an ongoing effort and new results are expected soon. The limits from a combined NUMI-MB + KARMEN + NOMAD analysis are shown on the middle panel of Fig.~\ref{fig2a}. Both panels illustrate that the null appearance experiments provide essentially no constraints to the parameter space allowed by the BNB-MB and LSND data sets, except at higher $\Delta m^2$. 

\indent The best-fit parameters obtained independently from the NUMI-MB and KARMEN data sets, shown in Table \ref{tab:3plus1fits}, are similar to those of LSND, BNB-MB($\nu$), and BNB-MB($\bar{\nu}$). The NOMAD data set, on the other hand, prefers a much larger $\Delta m_{41}^2\sim$50eV$^{2}$, and a much smaller $\sin^2 2\theta_{\mu e}\sim$10$^{-4}$. 

\indent A combined analysis of all appearance data yields a $\chi^2$-probability of 24\% for the best-fit hypothesis, both in the case where maximal mixing is allowed in the fit, and in the case where the electron and muon content of the sterile mass eigenstate has been limited to small values ($<$0.3). The allowed region obtained by a joint analysis of all appearance experiments under a (3+1) oscillation scenario is shown in the right panel of Fig.~\ref{fig2a}. 

\indent Similarly, Fig.~\ref{fig2b} (left) corresponds to the allowed region obtained by a joint analysis of BNB-MB($\bar{\nu}$) + LSND. The limit shown is that of the KARMEN experiment, which is the only other SBL experiment to perform an appearance search with antineutrinos. The KARMEN limit provides substantial coverage of the joint LSND and BNB-MB($\bar{\nu}$) allowed region, excluding the best-fit point of the LSND + BNB-MB($\bar{\nu}$) fit at $>$99\% CL. However, KARMEN imposes little constraint to the lower-$\Delta m^2$ allowed solutions. A joint analysis of all three data sets yields a $\chi^2$-probability of 29\% for the best-fit hypothesis, and an allowed region shown on the right panel of Fig.~\ref{fig2b}. The $\chi^2$-probability remains the same for fits where the electron and muon content of the sterile mass eigenstates have been limited to values less than 0.3. As shown in Table \ref{tab:3plus1fitsPG}, the three data sets are compatible at 15.3\%. New results from MiniBooNE with increased antineutrino statistics should be able to provide more information to these fits \cite{mbnubar}. 

\subsubsection{Studies with appearance and disappearance experiments}

\begin{figure}[htb]
\includegraphics[width=7cm, trim=10 10 10 10]{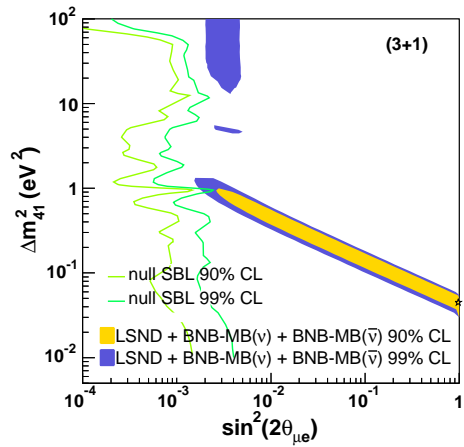} 
 \caption{\label{fig3} The allowed 90\% and 99\% CL regions (light and dark filled areas, respectively) from a combined analysis of BNB-MB($\nu$) + BNB-MB($\bar{\nu}$) + LSND data sets, and 90\% and 99\% exclusion limits (light and dark curves, respectively) from a combined analysis of all remaining (null, appearance and disappearance) SBL experiments. The null fit includes atmospheric constraints. The null SBL experiments exclude the joint 99\% CL allowed region at 99\% CL.}
\end{figure}

\indent Much stronger constraints than those of the null appearance experiments are provided by the null disappearance experiments (CCFR84, CDHS, CHOOZ, and Bugey) and atmospheric constraints, under the assumptions of CPT conservation and unitarity of the neutrino mixing matrix. The 90\% and 99\% CL exlusion limits from a combined analysis of all null data sets (NUMI-MB, KARMEN, NOMAD, Bugey, CHOOZ, CCFR84, CDHS, and atmospheric constraints) are shown in Fig.~\ref{fig3}. The figure shows that the parameter space jointly-allowed by BNB-MB($\nu$) + BNB-MB($\bar{\nu}$) + and LSND at 99\% CL is excluded by a combined analysis of all null SBL experiments, appearance and disappearance, at 99\% CL. The severe tension between LSND and the null SBL experiments \cite{Maltoni:2007zf} continues to exist and in fact increases further with the addition of BNB-MB($\nu$) and BNB-MB($\bar{\nu}$) data. The signal results show low (0.15\%) compatibility with null results. The LSND result remains to be mostly responsible for the low compatibility, as the BNB-MB($\nu$) and BNB-MB($\bar{\nu}$) experiments show 14\% and 3.7\% compatibility with the null experiments, respectively. 

\begin{figure*}[htb]
\vspace{0.5cm}
\includegraphics[width=7cm, trim=10 10 10 10]{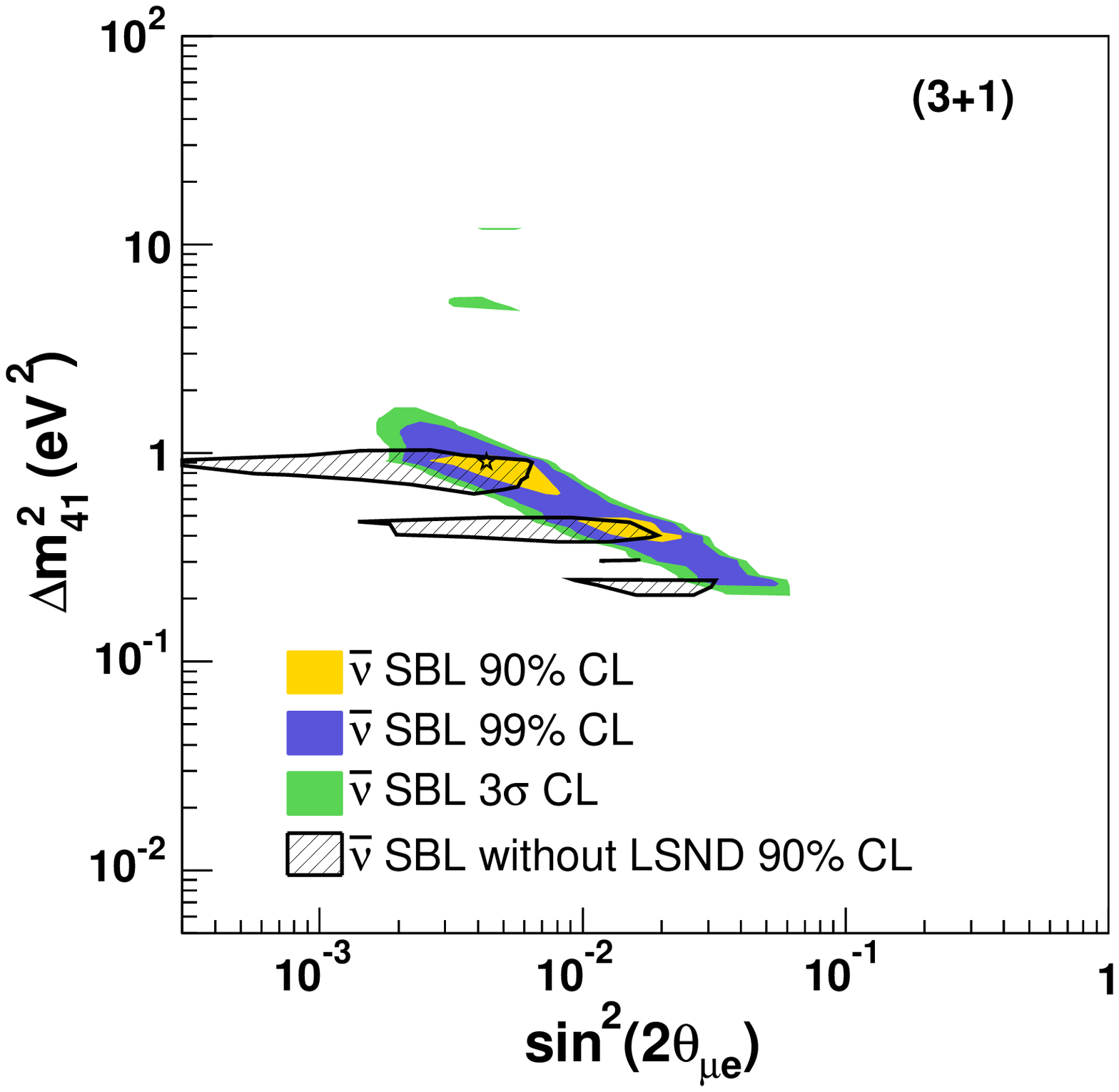}
\hspace{0.1cm}
\includegraphics[width=7cm, trim=10 10 10 10]{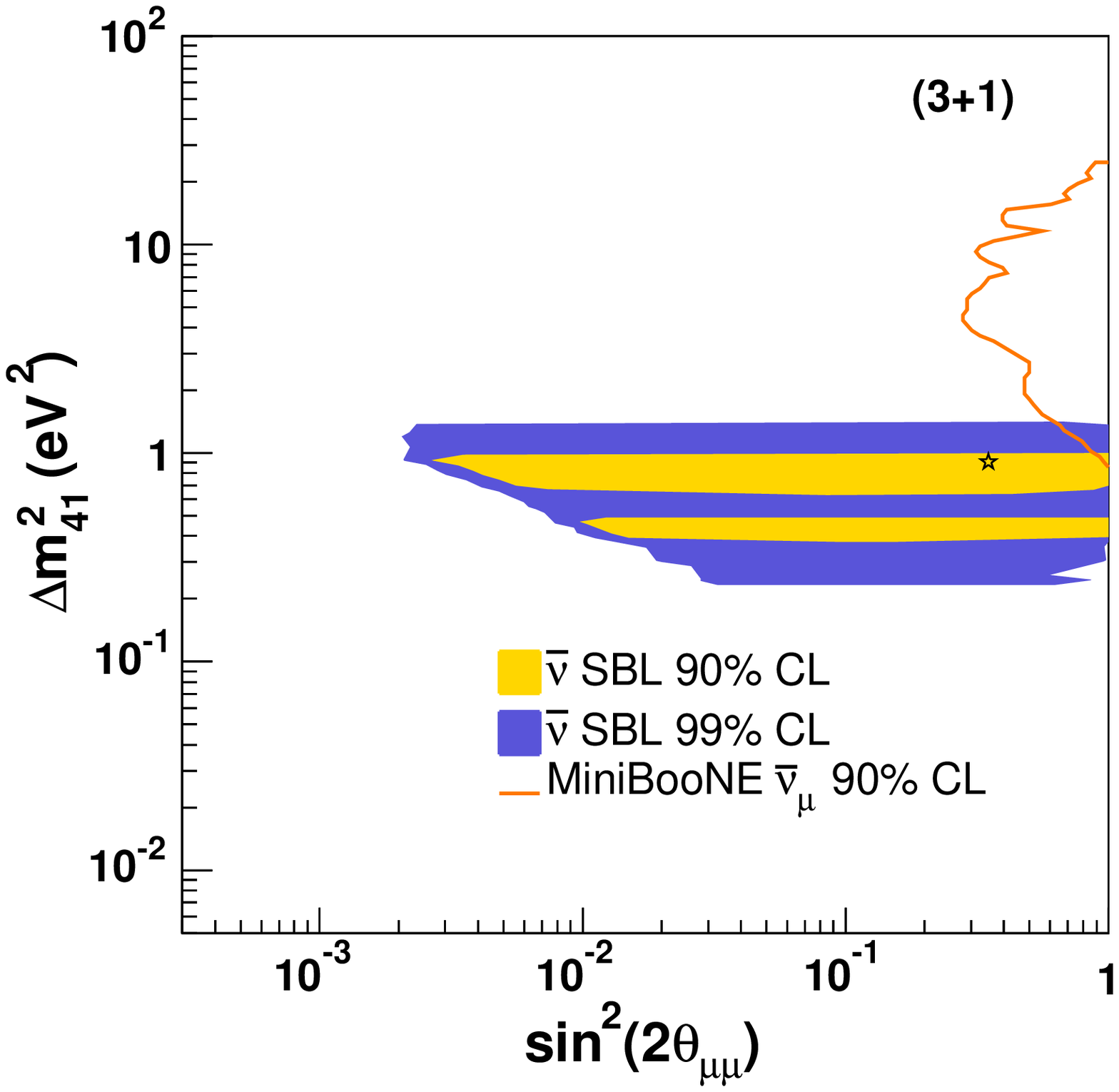}
\vspace{0.5cm}
 \caption{\label{fig3a} The allowed 90\%, 99\%, and 3$\sigma$ CL regions from a combined analysis of all antineutrino SBL data sets. The left plot also shows the 90\% CL allowed region obtained from a combined analysis of all antineutrino experiments except LSND (KARMEN, BNB-MB($\bar{\nu}$), Bugey, and CHOOZ). The right plot also shows the 90\% CL exclusion limit from \cite{mbnumudis}. The MiniBooNE $\bar{\nu}_{\mu}$ disappearance search excludes the parameter space to the right of the line at 90\% CL. See text for more details.}
\end{figure*}
\begin{figure*}[htb]
\vspace{0.5cm}
\includegraphics[width=7cm, trim=10 10 10 10]{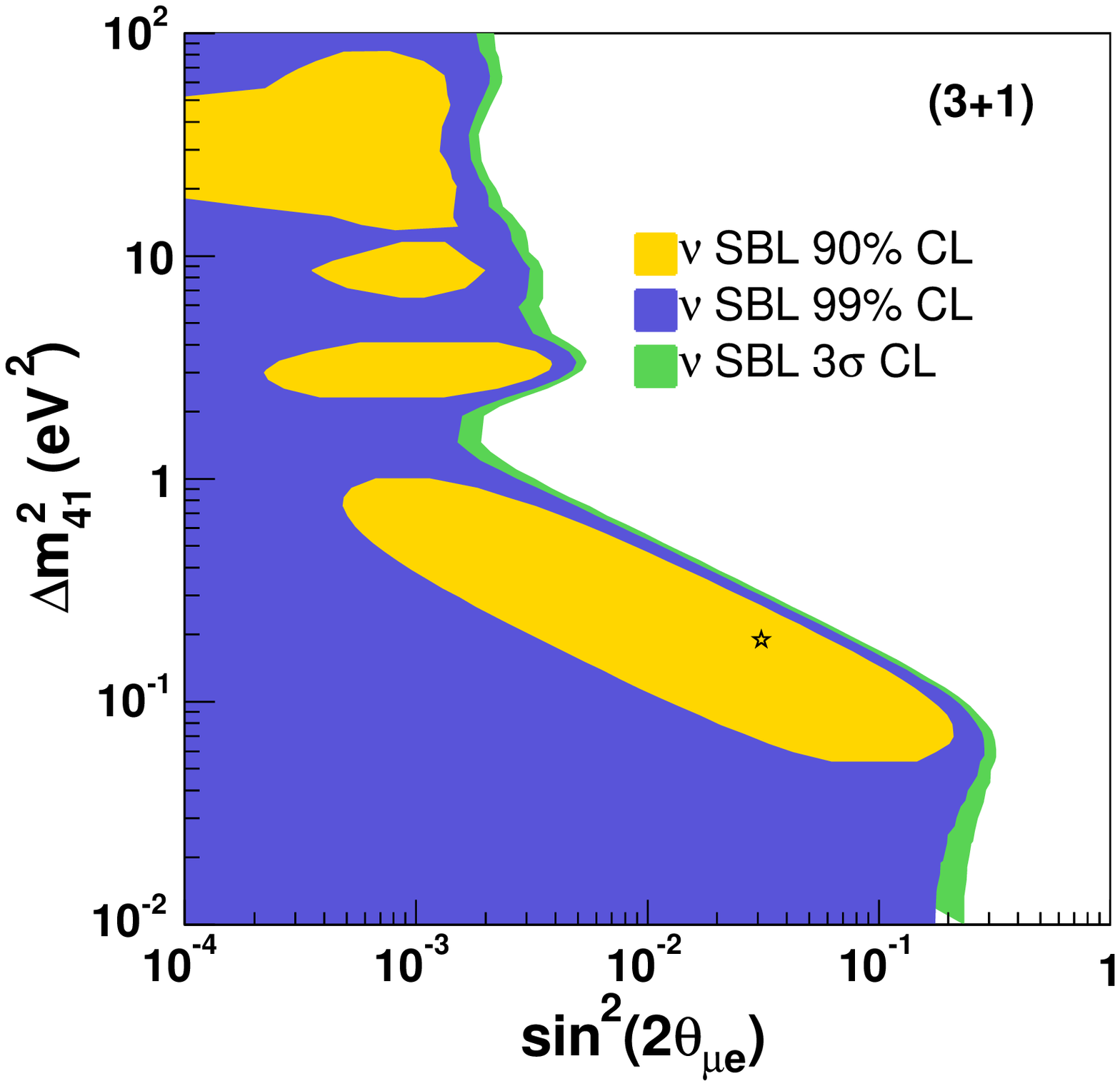} 
\hspace{0.1cm}
\includegraphics[width=7cm, trim=10 10 10 10]{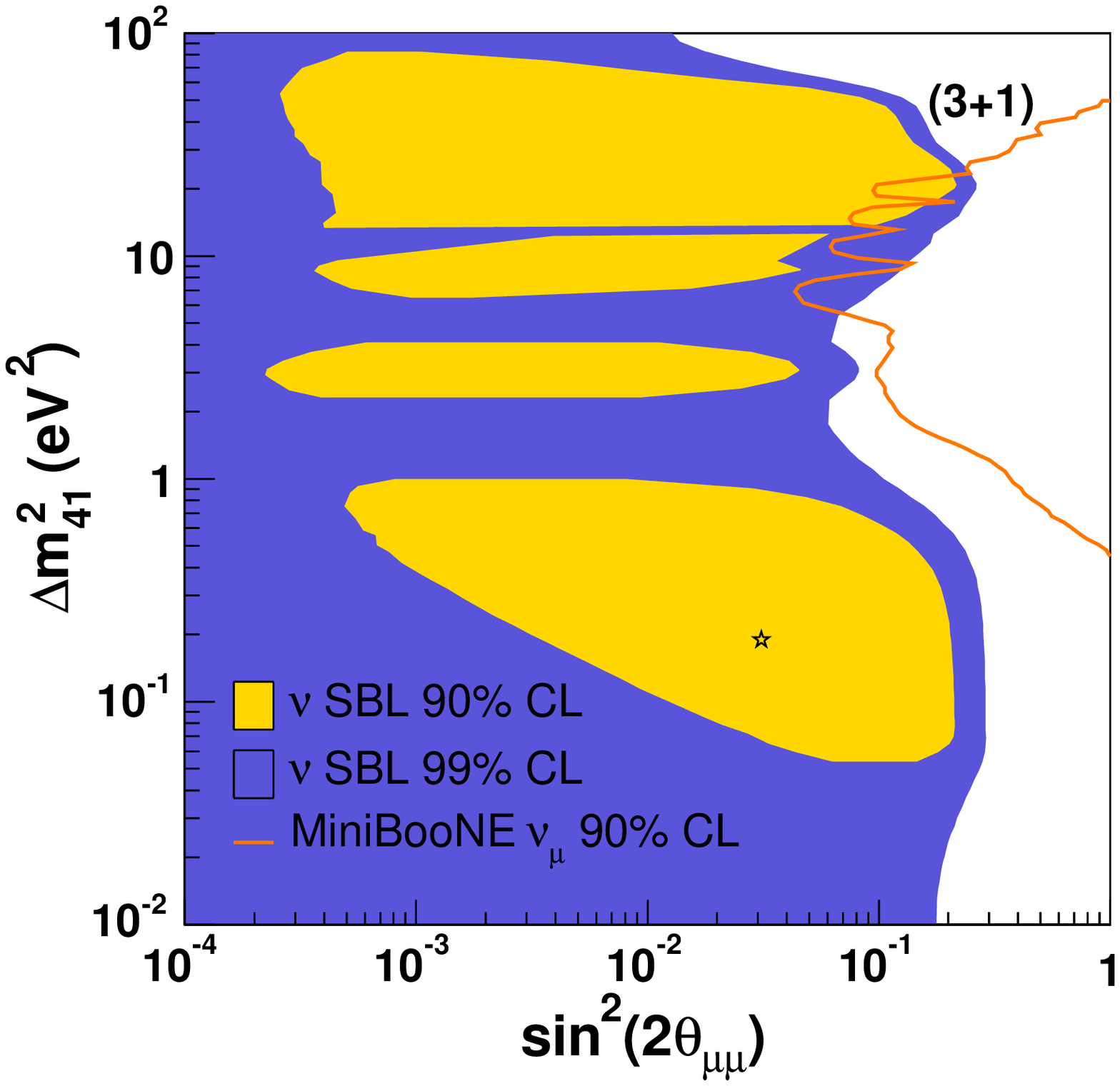}
\vspace{0.5cm}
 \caption{\label{fig3b} The allowed 90\%, 99\%, and 3$\sigma$ CL regions from a combined analysis of all neutrino SBL data sets. The right plot also shows the 90\% CL exclusion limit from \cite{mbnumudis}. The MiniBooNE $\nu_{\mu}$ disappearance search excludes the parameter space to the right of the line at 90\% CL. See text for more details.}
\end{figure*}

\indent Figure \ref{fig3a} shows the allowed region obtained by the joint BNB-MB($\bar{\nu}$) + LSND + KARMEN + Bugey + CHOOZ analysis. Here, the $\bar{\nu}_e$ disappearance constraints from Bugey and CHOOZ are interesting to consider from the perspective of a joint analysis of only antineutrino SBL experiments. In a joint fit, all of the above (antineutrino) experiments yield a high $\chi^2$-probability of 86\%, and 29.9\% compatibility. In these fits, Bugey and CHOOZ constrain $|U_{e4}|$, but provide no direct constraints on $|U_{\mu4}|$. However, a joint analysis with the LSND, BNB-MB($\bar{\nu}$), and KARMEN appearance experiments, which are sensitive to the product of $|U_{e4}||U_{\mu4}|$, provides indirect constraints to the $|U_{\mu4}|$ mixing element. Figure ~\ref{fig3a} (left) also shows that a fit to all antineutrino experiments without LSND yields similar closed contours at 90\% CL, which include the best-fit point. Current constraints from MiniBooNE on $\bar{\nu}_{\mu}$ disappearance \cite{mbnumudis} provide relatively small constraints to the $\sin^2 2\theta_{\mu\mu}$ allowed space, as illustrated in the right panel of Fig.~\ref{fig3a}. However, new results from a joint MiniBooNE and SciBooNE \cite{sciboone} $\bar{\nu}_{\mu}$ disappearance search, which are expected soon \cite{kendallprive}, may be able to probe this region with higher sensitivity, and will be interesting within the context of CPT-violating models. According to the best-fit oscillation parameters from a fit to only antineutrino SBL data, MiniBooNE should observe muon antineutrino disappearance with an amplitude of $\sin^2 2\theta_{\mu\mu}\sim$ 0.35, at $\Delta m^2\sim$ 0.91 eV$^2$. The MINOS experiment \cite{minos} should also have sensitivity to these oscillation parameters in antineutrino running mode; muon antineutrino disappearance search results from MINOS are expected soon \cite{wandc}. Incorporation of the upcoming MiniBooNE and MINOS disappearance results in these fits is currently being investigated. 

\indent Neutrino-only fits also yield a reasonably high $\chi^2$-probability of 47\%; the corresponding allowed regions are shown in Fig.~\ref{fig3b}. Current constraints from MiniBooNE $\nu_{\mu}$ disappearance are shown on the right panel of Fig.~\ref{fig3b}. Interestingly, fits to only neutrino SBL data also yield a closed contour at 90\% CL. The parameter space, however, points to smaller mixing amplitudes relative to those preferred by the antineutrino-only fit. Neutrino-only fits and antineutrino-only fits are incompatible, with a PG of 0.03\%, as shown in Table \ref{tab:3plus1fitsPG}. The large incompatibility between antineutrino and neutrino SBL results suggests that the neutrino and antineutrino data sets cannot be accommodated within a (3+1) CPT-conserving sterile neutrino oscillation scenario. However, the constraining power of antineutrino SBL experiments alone on $\Delta m^2_{41}$ and $\sin^2 2\theta_{\mu e}$ is remarkable, and invites exploration of models that provide the possibility of different oscillation patterns for neutrinos versus antineutrinos. 

\begin{figure*}[htbp]
\includegraphics[width=6cm,angle=-90]{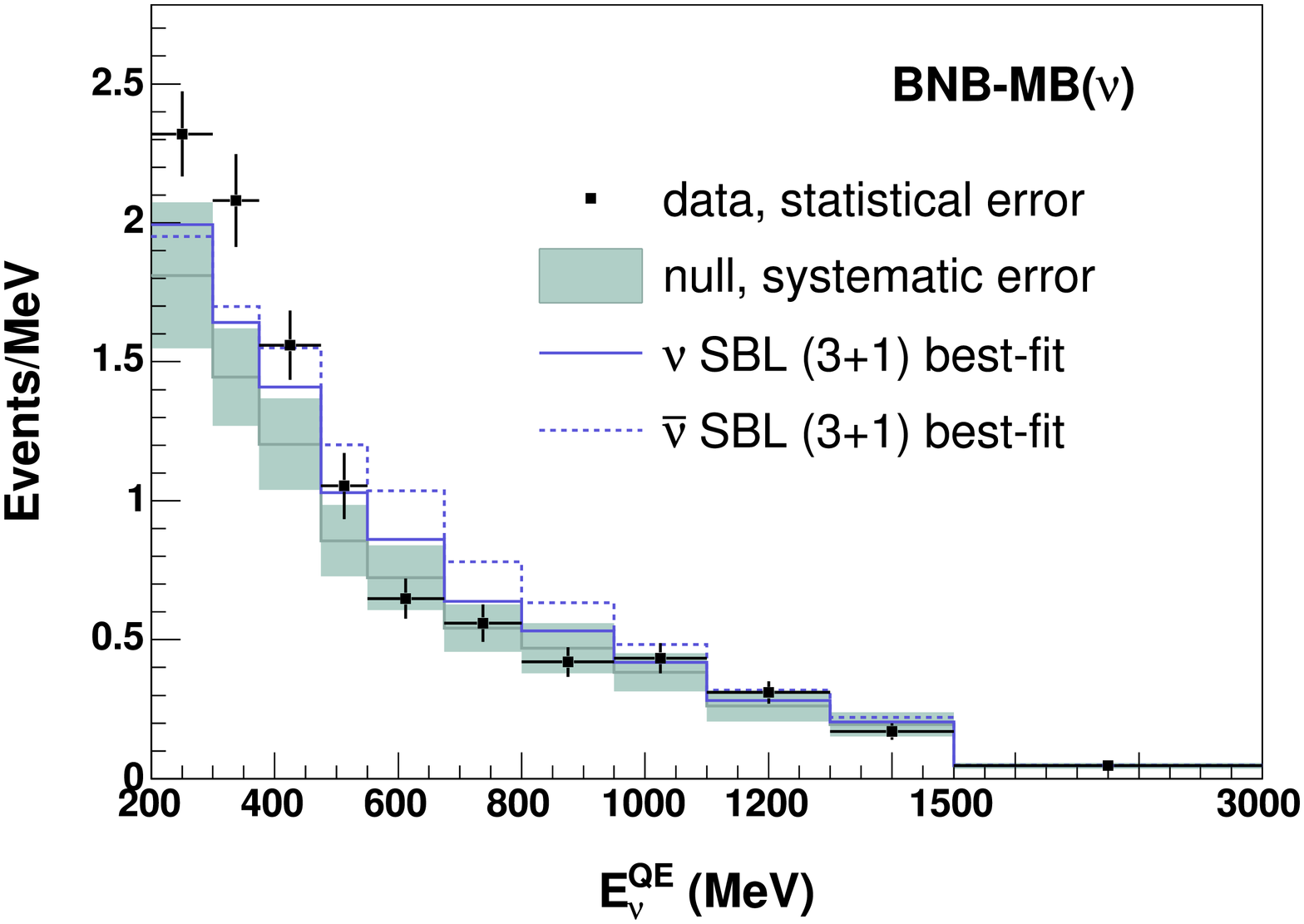}
\includegraphics[width=6cm,angle=-90]{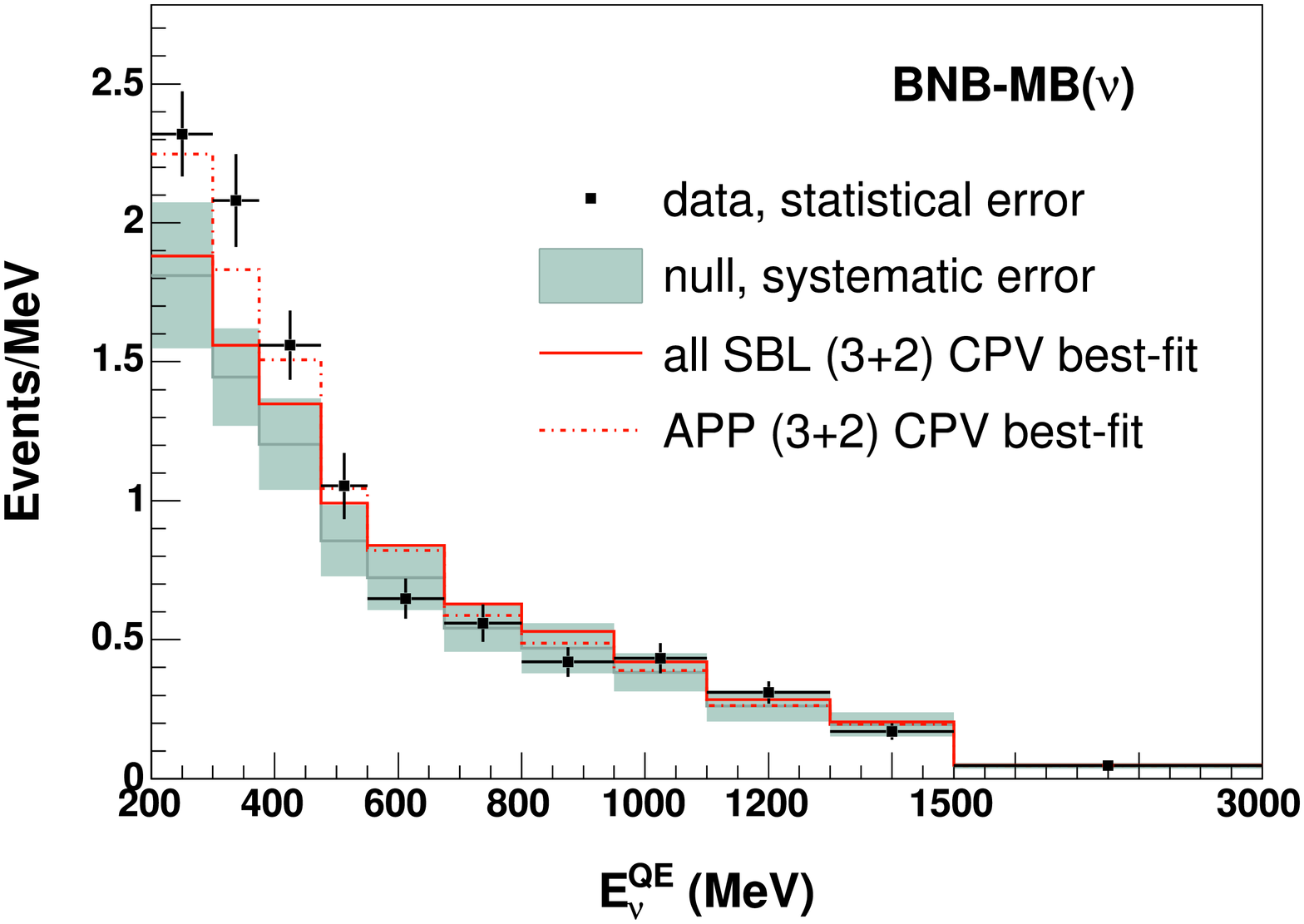}
\includegraphics[width=6cm,angle=-90]{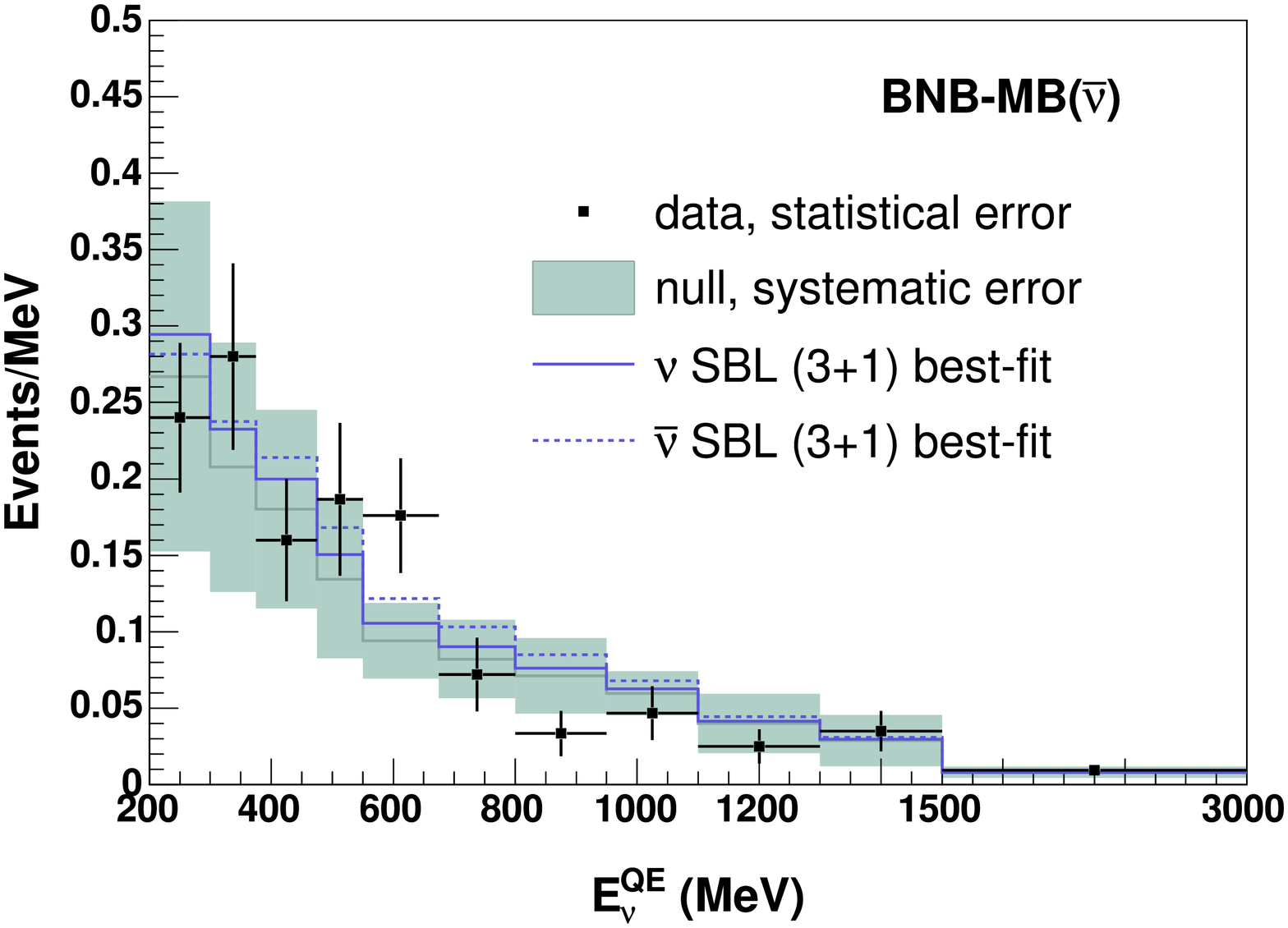}
\includegraphics[width=6cm,angle=-90]{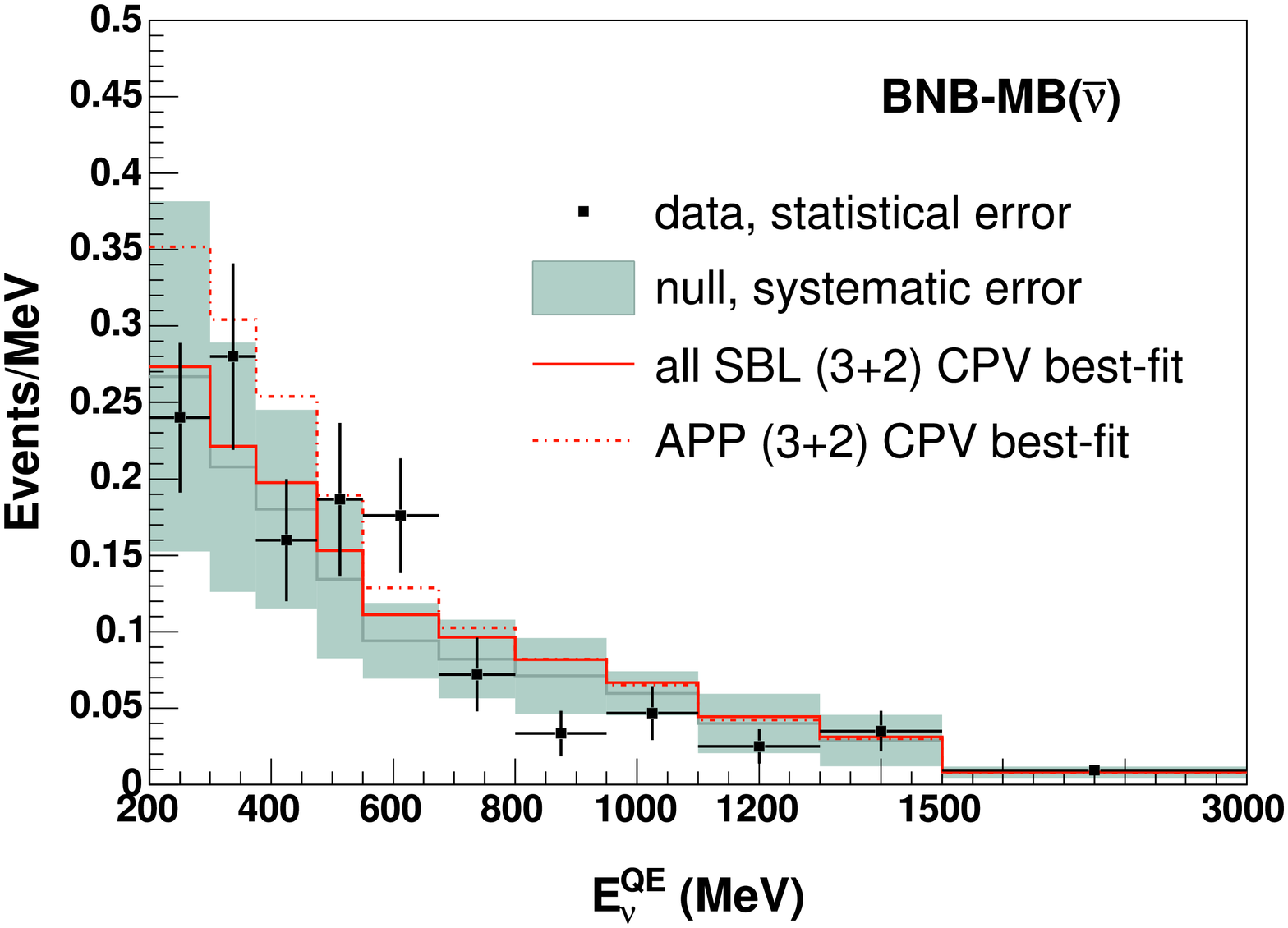}
\includegraphics[width=6cm,angle=-90]{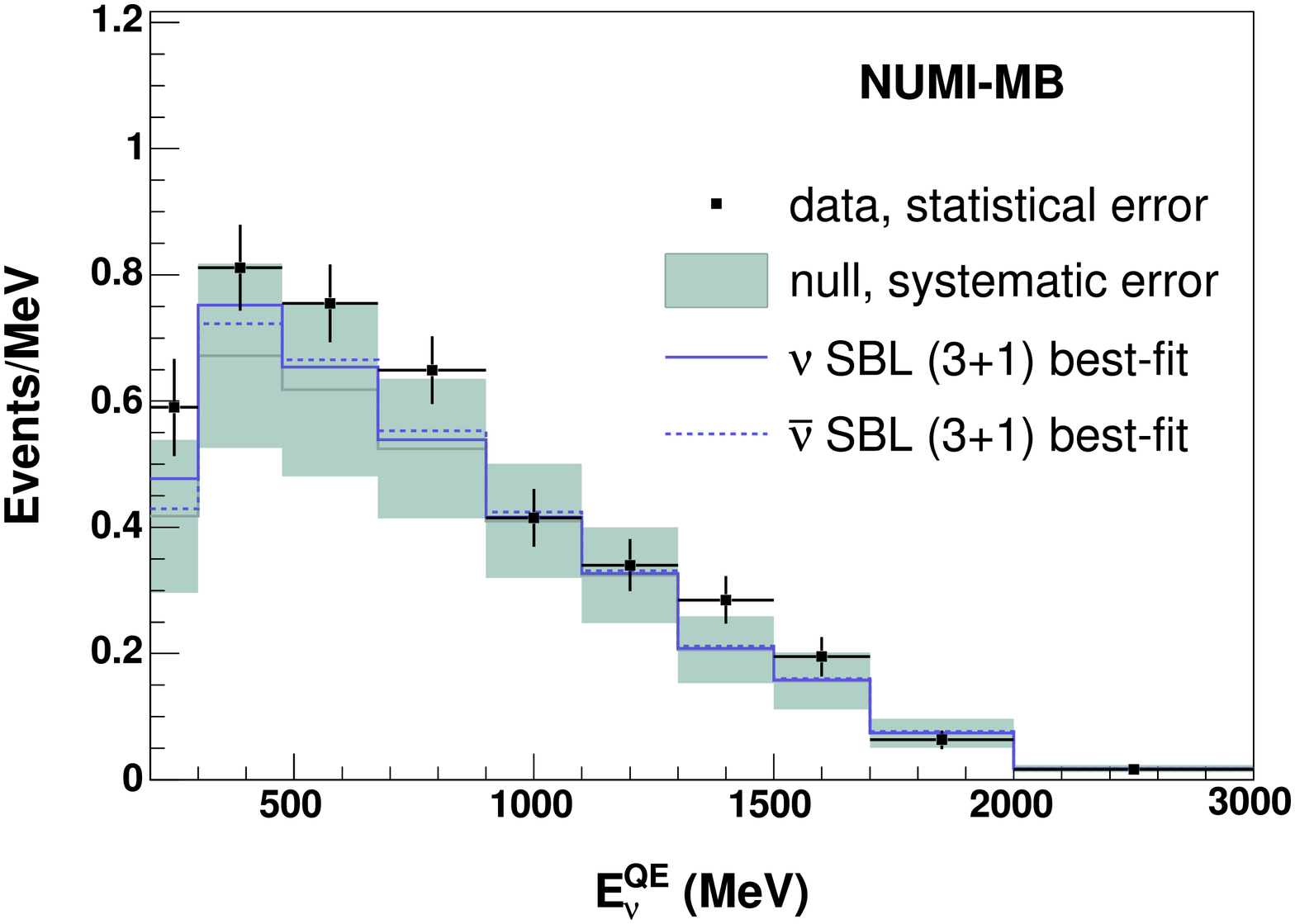}
\includegraphics[width=6cm,angle=-90]{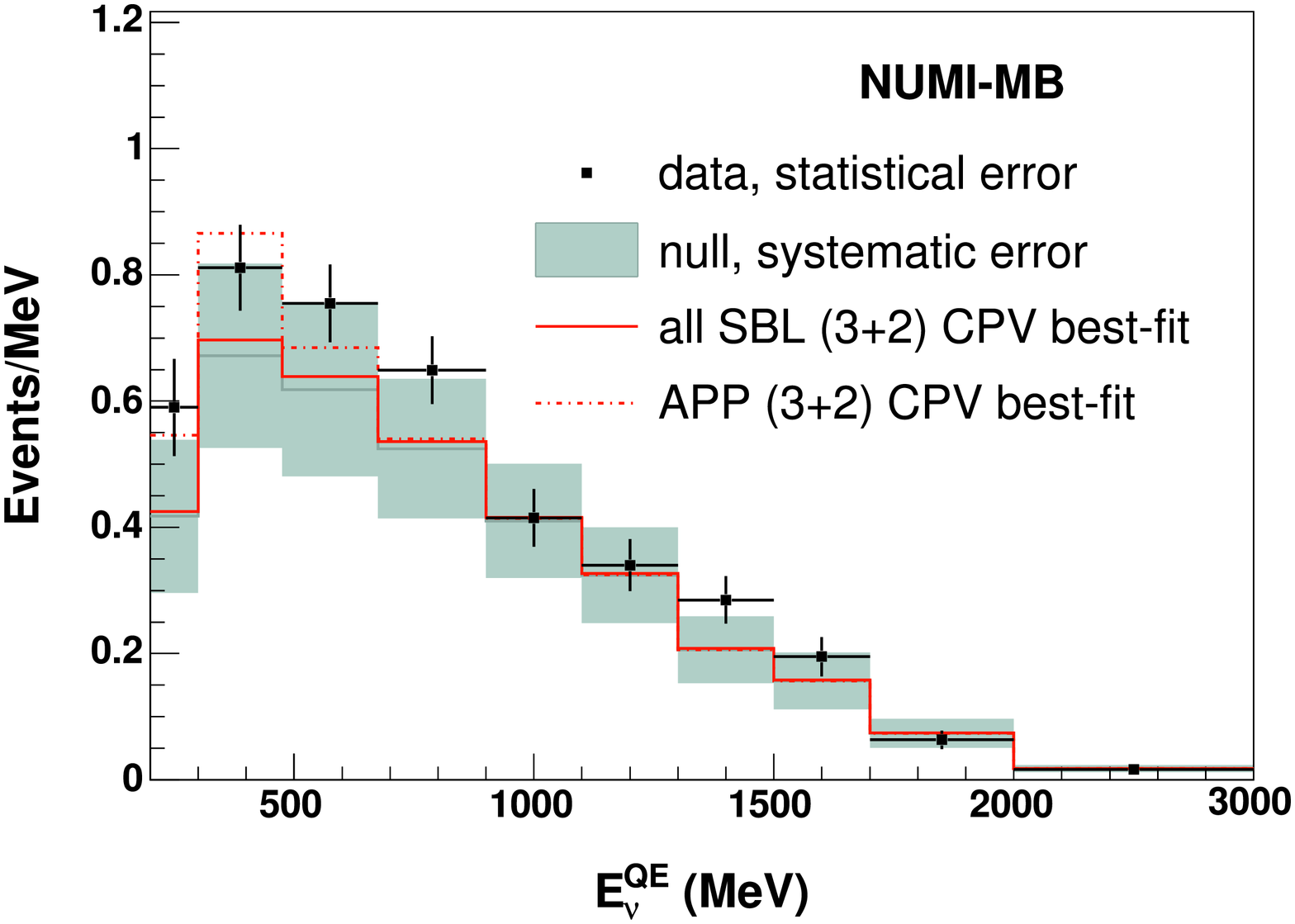}
\caption{\label{fig5pred} Left: MiniBooNE predicted event distributions using the neutrino-only (3+1) best-fit parameters ($\Delta m^2,\sin^2(2\theta)$)=(0.19, 0.031) in blue (dark gray) solid line and antineutrino-only (3+1) best-fit parameters ($\Delta m^2,\sin^2(2\theta)$)=(0.91, 0.0043) in blue (dark gray) dashed line. The null predictions are shown in light gray with systematic error bands. The observed data are shown in black points with statistical error bars. Right: MiniBooNE predicted event distributions using the best-fit parameters obtained from a (3+2) CP-violating fit to all SBL experiments and appearance-only SBL experiments, in red (dark gray) solid line and red (dark gray) dashed line, respectively.}
\end{figure*}

\begin{figure*}[tp]
\includegraphics[width=7.5cm, angle=-90, trim=150 10 20 0]{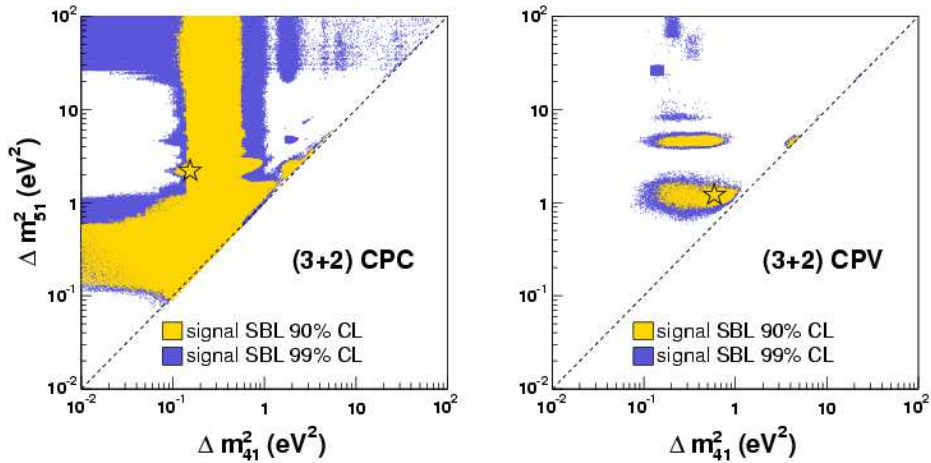} 
 \caption{\label{fig4} Allowed regions in ($\Delta m^2_{41}$,$\Delta m^2_{51}$) space for fits to CP-conserving (CPC, left) and CP-violating (CPV, right) (3+2) oscillation models. Only the BNB-MB($\nu$), BNB-MB($\bar{\nu}$) and LSND data sets have been included in the fit.}
\end{figure*}

\indent Figure \ref{fig5pred} (left) shows a comparison of the BNB-MB($\nu$), BNB-MB($\bar{\nu}$), and NUMI-MB event distributions for the neutrino-only best-fit parameters and antineutrino-only best-fit parameters. The neutrino best-fit parameters provide a better description to BNB-MB($\nu$) and NUMI-MB distributions than the antineutrino best-fit parameters, with $\chi^2_{BNB-MB(\nu)}=18.4$ and $\chi^2_{NUMI-MB}=4.4$, compared to $\chi^2_{BNB-MB(\nu)}=32.4$ and $\chi^2_{NUMI-MB}=4.8$. On the other hand, the antineutrino best-fit parameters provide a better description to BNB-MB($\bar{\nu}$) than the neutrino best-fit parameters ($\chi^2_{BNB-MB(\bar{\nu})}=19.7$, compared to $\chi^2_{BNB-MB(\bar{\nu})}=21.7$). 

\indent The best-fit results from the (3+1) oscillation fit involving all SBL data sets are summarized in Table \ref{tab:3plus1fits}. The best-fit parameters from neutrino-only and antineutrino-only fits are also shown. 

\subsection{(3+2) FIT RESULTS}

\indent Neutrino oscillation models with more than one sterile neutrino have been of particular interest because they open up the possibility of observable CP violation effects in short-baseline neutrino oscillations. If (3+$n$) sterile neutrino oscillations are realized in nature, with $n>$1, CP violation becomes a natural possibility, which is very appealing from the perspective of theories attempting to explain the matter-antimatter asymmetry in our universe \cite{Sakharov:1967dj}.

\indent In this section, the new MiniBooNE results are examined under both a CP-conserving (CPC) and a CP-violating (CPV) (3+2) oscillation hypothesis. The new results are studied first within the context of appearance-only experiments, and subsequently in fits involving both appearance and disappearance data. 

\indent From the point of view of the data at hand from LSND, BNB-MB($\nu$), and BNB-MB($\bar{\nu}$) (see Fig.~\ref{fig1}), CP violation offers the potential of reconciling two experimental signatures---an excess in LSND data at 3.8$\sigma$ significance and one suggested at 90\% CL in BNB-MB antineutrino data, both pointing to relatively large mixing, reconciled with a possible excess found in BNB-MB neutrino data suggesting relatively small mixing, both at a similar $L/E$---as manifestations of the same underlying oscillation hypothesis.

\begin{figure*}[tp]
\includegraphics[width=7.5cm, angle=-90, trim=150 10 20 0]{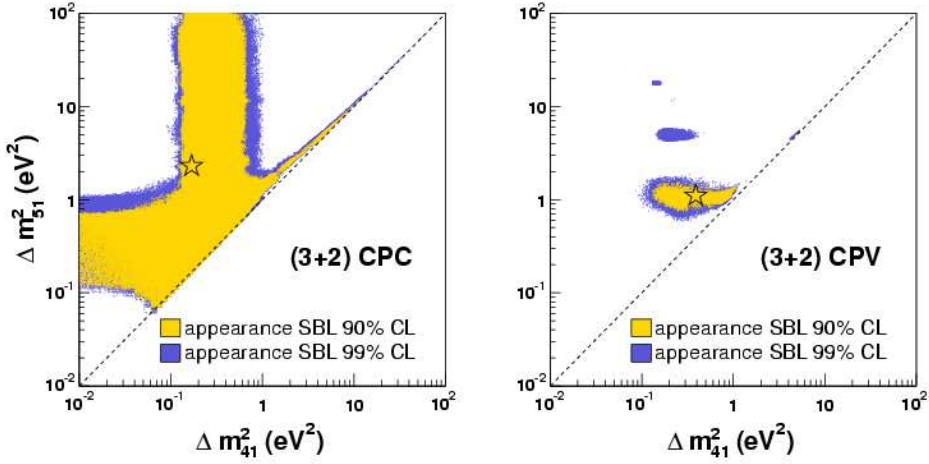}
 \caption{\label{fig5} Allowed regions in ($\Delta m^2_{41}$,$\Delta m^2_{51}$) space for fits to
 CP-conserving (CPC, left) and CP-violating (CPV, right) (3+2) oscillation models.
 Only appearance data sets have been included in the fit.}
\end{figure*}

\begin{figure*}[tp]
\includegraphics[width=7cm, angle=-90, trim=150 10 20 0]{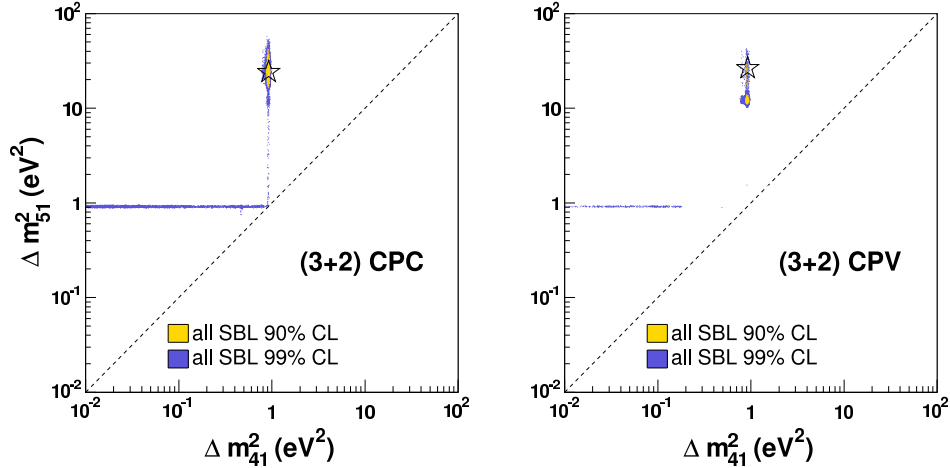} 
\vspace{0.5cm}
 \caption{\label{fig6} Allowed regions in ($\Delta m^2_{41}$,$\Delta m^2_{51}$) space for fits to CP-conserving 
 (CPC, left) and CP-violating (CPV, right) (3+2) oscillation models. All SBL data sets (appearance and disappearance) and atmospheric constraints have been included in the fit.}
\end{figure*}

\indent It should be noted that in the studies presented in this section, due to the larger dimensionality of the fits, the electron and muon content of the sterile mass eigenstates have been limited to values less than 0.3. This is a realistic assumption for sterile neutrino oscillation models.

\subsubsection{Studies with appearance-only experiments}

\indent Allowing for CP violation in (3+2) fits to LSND and BNB-MB($\nu$ and $\bar{\nu}$) data leads to a significant reduction in absolute $\chi^2$ of 12.2, for 1 degree of freedom (dof), corresponding to a best-fit CPV phase $\phi_{45}=$1.1$\pi$. The $\chi^2$-probability of the fit increases from 13\% in the CPC case to 53\% in the CPV case. The same test can be performed using all appearance data. In this case, allowing for CP violation leads to a reduction in $\chi^2$ of 13.3 for 1 dof, with a best-fit CPV phase $\phi_{45}=$1.1$\pi$. The $\chi^2$-probability from the CPV fit is comparable to that of a signal-only fit, at 56\%.

\indent The 90\% and 99\% CL allowed ($\Delta m^2_{41},\Delta m^2_{51}$) parameter space obtained by a combined fit to BNB-MB($\nu$) + BNB-MB($\bar{\nu}$) + LSND is shown in Fig.~\ref{fig4}. The figure illustrates that a CPV scenario (right panel) is much more restrictive in $\Delta m^2$ parameters compared to a CPC scenario (left panel). That is true both at 90\% and 99\% CL, shown by the significant reduction in allowed regions around $\Delta m^2_{41}=$0.5 eV$^2$ and $\Delta m^2_{51}=$1 eV$^2$.

\indent A similar effect is seen in the case of fits to all appearance experiments, as shown in Fig.~\ref{fig5}. Allowing for CP violation in fits to neutrino and antineutrino appearance data sets lead to a considerable improvement in the fit quality, and provides strong constraints to the $\Delta m^2_{41}$ and $\Delta m^2_{51}$ parameters of the model.

\indent The best-fit parameters for the signal-only and appearance-only fits are summarized in Table \ref{tab:3plus2bestfitresults}. 

\subsubsection{Studies with appearance and disappearance experiments}

\indent A dramatic reduction in the allowed ($\Delta m^2_{41},\Delta m^2_{51}$) parameter space occurs once all SBL data sets are considered in the fit, as shown in Fig.~\ref{fig6}. Compared to the CPC hypothesis, with the addition of disappearance constraints, the CPV hypothesis fails to provide a substantially better description of the data, reflected by the reduction in $\chi^2$ of $\chi^2_{CPC}-\chi^2_{CPV}=$2.2 for 1 dof. Furthermore, mainly due to CDHS \cite{Sorel:2003hf}, the allowed $\Delta m^2$ regions shift to higher $\Delta m^2_{51}$ values near $\Delta m^2_{51}=$25 eV$^2$.

\indent The returned $\chi^2$-probabilities from fits to all SBL data are 52\% for the CPC fit, and 54\% for the CPV fit. A PG test among all experimental data sets for the CPV case yields a compatibility of 7.0\%. While the $\chi^2$-probability and compatibility for the (3+2) CPV scenario are perfectly acceptable, as will be discussed in Sec.~\ref{sec:five}, an underlying source of tension exists due LSND and three other data sets: BNB-MB($\nu$), CDHS, and atmospheric constraints. The best-fit parameters extracted from a fit to all SBL data are also summarized in Table \ref{tab:3plus2bestfitresults}.

\begingroup
\squeezetable
\begin{table*}[htp]
\begin{ruledtabular} 
\begin{tabular}{ccccccccccr} 
\hline
Data Set & Fit & $\chi^2\ (dof)$ & $\chi^2$-probability & $\Delta m^2_{41}$ & $\Delta m^2_{51}$ & $|U_{e4}|$ & $|U_{\mu4}|$ & $|U_{e5}|$ & $|U_{\mu5}|$ & $\phi_{45}$ \\ \hline \hline
signal APP & CPV & 34.7(36)   & 53\% & 0.59 & 1.21 & 0.19 & 0.33 & 0.20  & 0.16 & 1.1$\pi$ \\   
signal APP & CPC & 46.9(37)   & 13\% & 2.01 & 2.22 & 0.42 & 0.24 & 0.33  & 0.33 & 0        \\   
APP        & CPV & 82.5(85)   & 56\% & 0.39 & 1.10 & 0.40 & 0.20 & 0.21 & 0.14  & 1.1$\pi$ \\	     
APP        & CPC & 95.8(86)   & 22\% & 0.18 & 2.31 & 0.32 & 0.38 & 0.086 & 0.071 & 0        \\
all SBL    & CPV & 189.3(192) & 54\% & 0.92 & 26.5 & 0.13 & 0.13 & 0.078 & 0.15 & 1.7$\pi$ \\
all SBL    & CPC & 191.5(193) & 52\% & 0.92 & 24.0 & 0.12 & 0.14 & 0.070 & 0.14 & 0 \\
\hline
\end{tabular} 
\end{ruledtabular} 
\caption{\label{tab:3plus2bestfitresults} Comparison of best-fit values for mass splittings and mixing parameters for (3+2) CP-conserving (CPC) and CP-violating (CPV) models. Mass splittings are shown in eV$^2$. The appearance experiments include BNB-MB($\nu$ and $\bar{\nu}$), LSND, NUMI-MB, KARMEN, and NOMAD. The signal experiments include LSND, BNB-MB($\nu$), and BNB-MB($\bar{\nu}$). See text for more details.} 
\end{table*} 
\endgroup

\begin{figure}[tbp] 
\vspace{-0.5cm}
 \includegraphics[ width=6.5cm, trim=70 20 0 0]{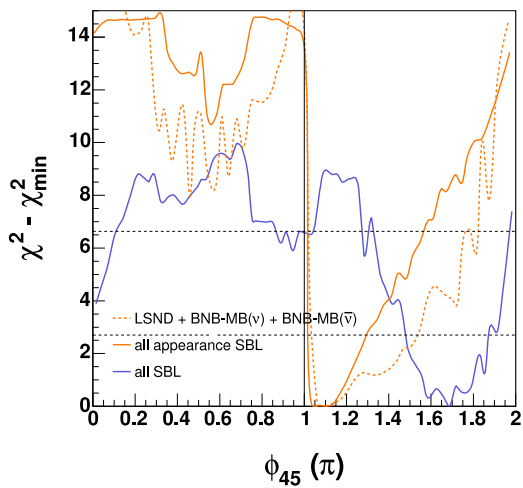} 
 \caption{\label{fig8} Projection of $\Delta\chi^2=\chi^2-\chi^2_{min}$ as a function of the CP-violating phase $\phi_{45}$. The dashed horizontal lines indicate the 90\% and 99\% CL $\Delta\chi^2$.}
\end{figure} 

\indent A comparison of Tables \ref{tab:3plus1fits} and \ref{tab:3plus2bestfitresults} suggests that, with the addition of the new data sets from MiniBooNE, the (3+2) CPV oscillation hypothesis provides a better description of all SBL data, compared to the (3+1) hypothesis. Compared to (3+1) models, (3+2) CP-conserving models give a reduction of 5.9 $\chi^2$ units for 3 additional fit parameters, while (3+2) CP-violating models give a reduction of 8.1 $\chi^2$ units with 4 additional parameters. This represents a relative improvement that is significantly smaller than that found in Ref.~\cite{Sorel:2003hf} from fits using data sets prior to atmospheric constraints and the new MiniBooNE results. 

\indent The MiniBooNE event distributions for the (3+2) CPV best-fit parameters are shown on the right panels of Fig.~\ref{fig5pred}. The resulting $\chi^2$-probabilities are 8.6\%, 6.7\%, and 33.6\%, for BNB-MB($\nu$), BNB-MB($\bar{\nu}$), and NUMI-MB, respectively, obtained using the best-fit parameters from a (3+2) CPV fit to all SBL data. Notice, however, that the best-fit parameters from a (3+2) CPV fit to appearance-only SBL data provide a better description of all three MiniBooNE data sets than the best-fit parameters from a (3+2) CPV fit to all SBL data, particularly for the BNB-MB($\nu$) data set. Furthermore, in the case of NUMI-MB, the (3+2) CPV appearance-only best-fit distribution, shown in dashed red (dark gray) on the right ($\chi^2$-probability$=$61\%), has comparable agreement with data as the (3+1) neutrino-only best-fit distribution, shown in solid blue (dark gray) on the left ($\chi^2$-probability$=$82\%). In the case of BNB-MB($\nu$), the (3+2) CPV appearance-only best-fit parameters are actually more preferred than the (3+1) neutrino-only best-fit parameters, with a $\chi^2$-probability of 56.9\%, rather than 30.1\%. However, in the case of BNB-MB($\bar{\nu}$) the $\chi^2$-probability is highest (23.4\%) for the (3+1) antineutrino-only best fit parameters.

\indent Figure \ref{fig8} shows the projection of $\Delta\chi^2=\chi^2-\chi^2_{min}$ as a function of the CP-violating phase $\phi_{45}$ for the three fits discussed in this section: the appearance-only fit projection is shown in the solid orange (light gray) line, the BNB-MB($\nu$)+BNB-MB($\bar{\nu}$)+LSND fit projection in dashed orange (light gray), and the projection from a fit to all SBL experiments is shown in blue (dark gray). Both the fit to the three signal experiments (BNB-MB($\nu$)+BNB-MB($\bar{\nu}$)+LSND) and the fit to appearance-only experiments seem to strongly prefer a CPV phase at $\phi_{45}=$1.1$\pi$, as illustrated by the three overlapping dips in the $\Delta\chi^2$ distribution. However, when fits to all SBL data are considered, the strong dependence disappears and a CPV phase at $\phi_{45}=$1.7$\pi$ is preferred.


\section{\label{sec:five}CONSTRAINTS TO (3+2) CP-VIOLATING FITS FROM EACH SBL EXPERIMENT}

In this section we study the constraints to experimentally allowed (3+2) CP-violating oscillations by each of the SBL experiments considered in our fits. This is accomplished through a study where fits are performed using all-but-one experiment at a time. Within this study, we are also interested in examining the source of incompatibility between appearance and disappearance data, as well as testing compatibility between neutrino and antineutrino appearance search results within a CP-violating scenario. The latter is motivated by the large incompatibility found in neutrino versus antineutrino fits, as well as appearance versus disappearance fits.

\begingroup
\squeezetable
\begin{table*}[htbp] 
\begin{ruledtabular} 
\begin{tabular}{l|cc|lcc} 
Data Set & $\chi^2\ (dof)$ & $\chi^2$-probability (\%) & \ \ \ \ \ \ \ \ \  PG (\%) & & \\ 

\hline \hline

all SBL   & 189.3 (192)    & 54.2 & PG( BNB-MB($\nu$),BNB-MB($\bar{\nu}$),NUMI-MB,LSND,KARMEN, &                    &   \\
                           &      &      & \ \ \ \ \ \ NOMAD,Bugey,CHOOZ,CCFR84,CDHS,ATM )    & = Prob( 53.9,40 ) = & 7.0 \\
          &                &      & PG( APP,DIS )                       & = Prob( 25.5,4 ) =                        & 0.004\\
          &                &      & PG( $\nu$,$\bar{\nu}$ )             &  = Prob( 25.4,7 ) =                       & 0.06\\
\hline
APP       & 82.5 (85)      & 55.7 & PG( BNB-MB($\nu$),BNB-MB($\bar{\nu}$),NUMI-MB,LSND,KARMEN, &   &  \\
          &                &      & \ \ \ \ \ \ NOMAD )                &  = Prob( 20.2,25 ) =               & 73.6 \\
DIS       & 81.3 (103)     & 94.4 & PG( Bugey,CHOOZ,CCFR84,CDHS,ATM )  & = Prob( 8.14,11 ) =                & 70.1 \\
\hline
$\nu$  & 81.3 (86)       & 62.4 & PG( BNB-MB($\nu$),NUMI-MB,NOMAD,CCFR84,CDHS,ATM ) & = Prob( 17.3,17 ) =   & 43.4 \\
$\bar{\nu}$ & 82.6 (99)  & 88.3 & PG( BNB-MB($\bar{\nu}$),KARMEN,LSND,Bugey,CHOOZ )   & = Prob( 11.2,16 ) = & 79.7 \\
\hline
$\nu$ APP  & 45.1 (53)       & 77.0 & PG( BNB-MB($\nu$),NUMI-MB,NOMAD )   & = Prob( 3.07,10 ) =             & 98.0 \\
$\bar{\nu}$ APP & 27.1 (27)  & 46.0 & PG( BNB-MB($\bar{\nu}$),KARMEN,LSND ) & = Prob( 6.88,10 ) =           & 73.7 \\
           &                 &      & PG( $\nu$ APP,$\bar{\nu}$ APP ) & = Prob( 10.3,5 ) =                  & 6.8  \\

\hline

all - BNB-MB($\nu$)         & 167.3 (174)     & 62.8 & PG( all SBL - BNB-MB($\nu$) , BNB-MB($\nu$) ) & = Prob( 15.7,5 ) =    & 0.78\\
all - BNB-MB($\bar{\nu}$)   & 167.4 (174)     & 62.6 & PG( all SBL - BNB-MB($\bar{\nu}$) , BNB-MB($\bar{\nu}$) ) & = Prob( 8.62,5 ) =  & 13 \\
all - NUMI-MB               & 183.7 (182)     & 45.1 & PG( all SBL - NUMI-MB , NUMI-MB ) & = Prob( 3.90,5 ) =   & 56 \\
all - LSND                  & 175.2 (187)     & 72.2 & PG( all SBL - LSND , LSND ) & = Prob( 12.5,5 ) =  & 2.9 \\
all - KARMEN                & 179.4 (183)     & 56.1 & PG( all SBL - KARMEN , KARMEN ) & = Prob( 4.53,5 ) =& 48 \\
all - NOMAD                 & 153.2 (162)     & 67.8 & PG( all SBL - NOMAD , NOMAD ) & = Prob( 1.96,5 ) = & 86 \\
all - Bugey                 & 140.4 (132)     & 29.2 & PG( all SBL - Bugey , Bugey ) & = Prob( 3.90,4 ) = & 42 \\
all - CHOOZ                 & 179.9 (178)     & 44.6 & PG( all SBL - CHOOZ , CHOOZ ) & = Prob( 3.09,4 ) = & 54 \\
all - CCFR84                & 174.3 (174)     & 47.9 & PG( all SBL - CCFR84 , CCFR84 ) & = Prob( 0.35,4 ) =  & 99 \\
all - CDHS                  & 172.8 (177)     & 57.5 & PG( all SBL- CDHS , CDHS ) & = Prob( 9.21,4 ) =  & 5.6\\
all - ATM                   & 184.0 (190)     & 60.9 & PG( all SBL - ATM , ATM ) & = Prob( 5.31,1 ) =   & 2.1\\
\end{tabular} 
\end{ruledtabular} 
\caption{\label{tab:tension} Comparison of $\chi^2$-probabilities for (3+2) CP-violating fits with different combinations of SBL data sets. Also shown are PG results testing compatibility among different data sets. The last eleven rows of the table provide the compatibility (PG) between the experiment being removed from each fit and all remaining experiments. See text for more details.}
\end{table*} 
\endgroup

\begingroup
\squeezetable
\begin{table}[htbp] 
\begin{ruledtabular} 
\begin{tabular}{l|c} 
Data Sets                                       & PG (\%) \\ \hline \hline
APP vs. DIS                                    & 0.004   \\
APP (no BNB-MB($\nu$)) vs. DIS (no CDHS + ATM) & 23.7   \\
APP (no BNB-MB($\nu$)) vs. DIS (no CDHS)       & 0.36   \\
APP (no BNB-MB($\nu$)) vs. DIS (no ATM)        & 0.52   \\
APP (no BNB-MB($\nu$)) vs. DIS                 & 0.067   \\
APP vs. DIS (no CDHS + ATM)                    & 2.9   \\
APP vs. DIS (no CDHS)                          & 0.027   \\
APP vs. DIS (no ATM)                           & 0.019    \\
\end{tabular} 
\end{ruledtabular} 
\caption{\label{tab:tension2} Comparison of compatibility between appearance (APP) and disappearance (DIS) experiments, within a (3+2) CP-violating scenario. The BNB-MB($\nu$) data set, CDHS data set, and atmospheric constraints (ATM) are removed from the fits as specified in order to establish the source of tension between appearance and disappearance experiments. Compatibilities are obtained using $ndf_{PG}=$4. See text for more details.}
\end{table} 
\endgroup

\begingroup
\squeezetable
\begin{table}[htbp] 
\begin{ruledtabular} 
\begin{tabular}{l|c} 
Data Sets                                       & PG (\%) \\ \hline \hline
$\nu$ vs. $\bar{\nu}$                                    & 0.06 \\
$\nu$ (no BNB-MB($\nu$) + CDHS + ATM) vs. $\bar{\nu}$    & 56.5 \\ 
$\nu$ (no BNB-MB($\nu$) + CDHS) vs. $\bar{\nu}$          & 3.7 \\ 
$\nu$ (no BNB-MB($\nu$) + ATM) vs. $\bar{\nu}$           & 4.4 \\ 
$\nu$ (no BNB-MB($\nu$)) vs. $\bar{\nu}$                 & 1.1 \\ 
$\nu$ (no CDHS + ATM) vs. $\bar{\nu}$                    & 2.3 \\ 
$\nu$ (no CDHS) vs. $\bar{\nu}$                          & 0.07 \\ 
$\nu$ (no ATM) vs. $\bar{\nu}$                           & 0.21 \\ 
\end{tabular} 
\end{ruledtabular} 
\caption{\label{tab:tension3} Comparison of compatibility between neutrino ($\nu$) and antineutrino ($\bar{\nu}$) experiments, within a (3+2) CP-violating scenario. The BNB-MB($\nu$) data set, CDHS data set, and atmospheric constraints (ATM) are removed from the fits as specified in order to establish the source of tension between neutrino and antineutrino experiments. Compatibilities are obtained using $ndf_{PG}=$7. See text for more details.}
\end{table} 
\endgroup

\indent Table \ref{tab:tension} summarizes the $\chi^2$-probability and PG results from (3+2) CP-violating fits. The upper rows summarize $\chi^2$-probabilities and PG's from fits to all SBL experiments, as well as fits to appearance-only, disappearance-only, neutrino-only, antineutrino-only, neutrino appearance-only, and antineutrino appearance-only data sets. Appearance and disappearance data sets, as well as neutrino and antineutrino data sets, are incompatible with a PG of less than 0.1\%. Grouping SBL appearance-only data sets according to whether they are neutrino or antineutrino experiments yields significantly higher compatibilities---98\% among $\nu$ appearance experiments, and 74\% among $\bar{\nu}$ appearance experiments. The compatibility between $\nu$ and $\bar{\nu}$ appearance-only results is lower, at 6.8\% but still acceptable. In the case where disappearance experiments are included in the comparison between neutrino and antineutrino fits, the compatibility among all $\bar{\nu}$ SBL data sets remains considerably high, at 80\%, as does the compatibility among all $\nu$ SBL data sets, at 43\%. However the compatibility between $\bar{\nu}$ and $\nu$ results is only 0.06\%.

\indent The remaining rows of Table \ref{tab:tension} provide the $\chi^2$-probabilities of global fits under the same oscillation scenario where one experiment is excluded from the fit at a time (as indicated by the ``-'' sign in the table). The $\chi^2$ probabilities of all fits are acceptable, ranging between 29.2\% for a fit excluding the Bugey data set, and 72.2\% for a fit excluding the LSND data set. Aside from LSND, three experiments stand out as having the poorest compatibility when compared to a global fit with all other SBL data sets: 1) BNB-MB($\nu$), 2) CDHS and 3) atmospheric constraints (ATM). These three experiments have been identified as the possible source of tension between appearance and disappearance experiments, or neutrino and antineutrino experiments. The remaining combinations yield reasonably high compatibilities of at least 42\%, with the exception of LSND and BNB-MB($\bar{\nu}$) which are compatible with the remaining data sets at 2.9\% and 13\%, respectively. 

\indent To further test the hypothesis that the tension between appearance and disappearance experiments is a result of the BNB-MB($\nu$) and CDHS data sets and atmospheric constraints, the compatibility between appearance and disappearance experiments is re-evaluated several times. Each time, a different combination of these three experiments is excluded from the fits. The results are summarized in Table \ref{tab:tension2}. The compatibility among appearance and disappearance experiments with BNB-MB($\nu$), CDHS, and atmospheric constraints excluded from the fits is high, at 23.7\%. The BNB-MB($\nu$) data set alone is not responsible for the disagreement between appearance and disappearance experiments, as suggested by the sixth row of the Table \ref{tab:tension2}. Even with BNB-MB($\nu$) included in the fit, a compatibility of 2.9\% can be obtained if CDHS and atmospheric constraints are excluded from the fit. 

\indent The same test can be performed between neutrino and antineutrino experiments. The results are summarized in Table \ref{tab:tension3}. Again, the compatibility between neutrino and antineutrino experiments is re-evaluated several times; each time, a different combination of the BNB-MB($\nu$), CDHS, and atmospheric constraint data sets is excluded from the fits. Here, the compatibility among neutrino and antineutrino experiments with BNB-MB($\nu$), CDHS, and atmospheric constraints excluded from the fits is even higher, at 56\%. The BNB-MB($\nu$) data set is just as responsible for the disagreement between neutrino and antineutrino experiments as the CDHS data set and atmospheric constraints alone. The tension seems to be caused by all three experiments, as none of them independently excluded from the fit can accound for the increase in compatibility from $\sim$1\% (or less) to 56\%.

\indent It is possible that higher compatibility between BNB-MB($\nu$) and all remaining SBL data sets may be achieved if the fits are to be repeated with the low energy region (200$<E^{QE}_{\nu}<$475 MeV) excluded from the BNB-MB($\nu$) data set. 

\indent A global analysis with BNB-MB($\nu$), CDHS, and atmospheric constraints excluded from the fit yields a $\chi^2$-probability of 82\% and $>$90\% compatibility among all experiments. 

\section{\label{sec:six}CONCLUSIONS}

We have re-examined global fits to sterile neutrino oscillation models, using new data from MiniBooNE. Those include the final MiniBooNE neutrino mode results and the first, low statistics MiniBooNE antineutrino results, as well as first results from the off-axis NuMI beam observed in the MiniBooNE detector. 

Within a (3+1) CP- and CPT-conserving scenario, we have found that the data set collected by MiniBooNE using the NuMI off-axis beam (NUMI-MB) currently provides very weak constraints to sterile neutrino fits, due to large systematic uncertainties. Updated NuMI results, expected soon, should have a greater impact on these fits. 

Within the same oscillation framework, the MiniBooNE antineutrino data set (BNB-MB($\bar{\nu}$)) is found in agreement with LSND, yielding, in a combined analysis with LSND and KARMEN under a (3+1) oscillation hypothesis, a $\chi^2$-probability of 29\%, and best-fit parameters similar to those of LSND. Updated MiniBooNE antineutrino appearance results, with almost twice the current statistics, are expected in the near future. 

The MiniBooNE neutrino data set (BNB-MB($\nu$)), although suggestive of an excess that could be described by a (3+1) oscillation hypothesis with a $\chi^2$ probability of 35\%, is found incompatible with the signals from the MiniBooNE antineutrino and LSND results. 

The remaining null appearance and disappearance experiments (NUMI-MB($\nu$), KARMEN, NOMAD, Bugey, CHOOZ, CDHS, CCFR84) and atmospheric oscillation data impose strong constraints to the parameter space allowed by a combined (3+1) analysis of MiniBooNE neutrino and antineutrino and LSND data, excluding the 99\% CL allowed region at 99\% CL. However, the constraints from antineutrino disappearance experiments on the parameter space allowed by antineutrino appearance experiments (BNB-MB($\bar{\nu}$), LSND, and KARMEN) are weaker. In a (3+1) oscillation framework, all antineutrino experiments yield a best-fit $\chi^2$-probability of 86\%, and exclude the no-oscillations hypothesis at $>$5.0$\sigma$. The best-fit parameters are similar to those of LSND, and correspond to a muon antineutrino disappearance amplitude of 0.35, which may be addressed by upcoming results from MiniBooNE and MINOS on muon antineutrino disappearance. Additionally, fits to all neutrino experiments yield a best-fit $\chi^2$-probability of 47\% and exclude the null hypothesis at $>$90\% CL.

Furthermore, we find that with the addition of the new MiniBooNE data sets, the (3+2) oscillation models provide a much better description of all SBL data sets compared to (3+1) models. In the case of (3+2) fits, CP violation allows for a significant improvement in $\chi^2$-probability for fits to only BNB-MB($\nu$) + BNB-MB($\bar{\nu}$) + LSND, and fits to only appearance experiments. In the case of global fits, however, the effect of CP violation is muted, as allowing for CP violation results in a relatively small improvement in the fit. The $\chi^2$-probability for the best-fit (3+2) CPV hypothesis is 54\%, compared to 52\% for the CPC case. The best-fit corresponds to large but not maximal CP violation ($\phi_{45}=1.7\pi$). 

The high incompatibility among appearance and disappearance data seen in the past \cite{Maltoni:2007zf} in the case of (3+2) CP-violating fits still remains with the addition of the new MiniBooNE results. We have shown that the incompatibility is a result of the BNB-MB($\nu$) and CDHS data sets and atmospheric constraints. The compatibility between appearance and disappearance data with these three experiments excluded from the fits is significantly higher, at 24\%. 

Neutrino and antineutrino results are also incompatible within a (3+2) CP-violating scenario, with a PG of less than 0.1\%. The compatibility improves to 6.8\% in the case of comparing appearance-only neutrino versus antineutrino results. 

Overall, allowing for mixing with multiple sterile neutrino states and CP violation does not seem sufficient to allow incorporating all SBL experiments within a CPT-conserving, sterile neutrino oscillation framework. It may be that there is an issue with one or more of the following data sets: LSND, BNB-MB($\nu$), including the low-energy excess, CDHS, or atmospheric constraints; alternatively, theories with CPT-violating oscillations or effective CPT violation \cite{barenboim,Barenboim:2009ts,pas} may succeed in reconciling all short-baseline oscillation signatures, and should be explored.


\begin{acknowledgments}
\noindent We thank Bill Louis for valuable discussions. We also thank the National Science Foundation for their support.
\end{acknowledgments}

%

\end{document}